\documentclass[twocolumn,trackchanges]{aastex7}

\newcommand\msun{M_\odot}

\usepackage{color}

\shorttitle{Sextans SFH}
\shortauthors{Andr\'es E. Piatti}

\begin{document}

\title{Spatially resolved star formation history of Sextans dSph}

\author[0000-0002-8679-0589]{Andr\'es E. Piatti}
\affiliation{Instituto Interdisciplinario de Ciencias B\'asicas (ICB), CONICET-UNCuyo, Padre J. Contreras 1300, M5502JMA, Mendoza, Argentina}
\affiliation{Consejo Nacional de Investigaciones Cient\'{\i}ficas y T\'ecnicas (CONICET), Godoy Cruz 2290, C1425FQB,  Buenos Aires, Argentina}
\email{e-mail: andres.piatti@fcen.uncu.edu.ar}

%\correspondingauthor{Andr\'es E. Piatti}

\begin{abstract}
We present a spatially resolved archaeological reconstruction of the Sextans 
dwarf spheroidal (dSph) galaxy using deep DECam wide-field photometry and
 the \texttt{PANCAKE} CMD-fitting code. By analyzing the star formation history 
 (SFH) and age-metallicity relationship (AMR) across four radial zones, namely:
 core, ring, outer body, and outskirts, we find that Sextans is a composite system 
 formed through a minor merger approximately 13 Gyr ago. Our results reveal 
 an inverse metallicity gradient: a primitive, metal-poor host (the current core) 
 surrounded by a more massive, chemically evolved envelope ($\Delta$[Fe/H] 
 $\approx$ -0.5 dex) introduced by the accreted satellite. We identify a distinct 
 delayed onset of star formation in the ring at $\sim$13 Gyr, marking the merger 
 event. While the core quenched early, star formation in the outer body and ring 
 persisted until $\sim$9 Gyr, suggesting that the final cessation of activity was 
 driven by environmental stripping during infall into the Milky Way halo.
 We propose a plausible scenario to reconcile the derived inverse metallicity 
gradient with the observed horizontal-branch (HB) morphology and reported Mg 
deficits. We suggest that the red HB dominance in the core reflects its ancient, 
$\alpha$-rich nature, while the blue HB in the outskirts represents an $\alpha$-poor, 
accreted component. However, we note that our CMD-derived [Fe/H] values are 
model-dependent 
inferences based on the total metal content $Z$. These findings suggest a 
non-monolithic assembly for Sextans, posing a testable prediction of a strong 
radial gradient in [$\alpha$/Fe].
 \end{abstract}


\section{Introduction}

The Sextans dwarf spheroidal (dSph) galaxy, discovered by \citet{irwinetal1990} via 
automated plate scans, stands as one of the most intriguing classical satellites 
of the Milky Way. Located at a distance of approximately 86$\pm$4 kpc 
\citep{karachentsevetal2004,mateo1998}, Sextans is characterized by a high degree 
of spatial extension and a notably low surface brightness ($\Sigma _V \approx$ 26.2 
mag arcsec$^{-2}$). 
Structurally, it exhibits a significant ellipticity \citep[$e$ $\approx$ 0.27;][]{ih1995}
and a large core radius ($r_c$ $\approx$ 13.8$\arcmin$ - 15$\arcmin$).
Historically, the galaxy was modeled with standard \citet{king62} or 
\citet{sersic1968} profiles under the assumption of dynamic equilibrium.

Early studies identified Sextans as a predominantly ancient system, with a star 
formation history (SFH) that truncated abruptly roughly 10-12 Gyr ago, likely due to 
environmental quenching from cosmic reionization or early tidal stripping 
\citep{mateoetal1991}. This fossil paradigm was reinforced by \citet{bettinellietal2018}, 
who utilized deep $B,I$ photometry from Subaru/Suprime-Cam to characterize Sextans 
as a 
reionization fossil that ceased star formation approximately 1.3 Gyr after the Big Bang. 
While that work found no significant metallicity gradient within the core, its focus remained 
largely restricted to the inner regions.
Its global metallicity is low, with a mean [Fe/H] $\approx$ -1.9 dex
\citep{battagliaetal2011,kirbyetal2011}, typical for its luminosity ($M_V \approx$ -9.3 mag). 
Its tidal radius, estimated at $r_t \approx$ 120\arcmin, 
suggests a system that has been significantly affected by the Milky Way’s potential, yet its
velocity dispersion \citep[$\sigma \approx$ 7.9 km/s;][]{walkeretal2009}
indicates a substantial dark matter halo with a high mass-to-light ratio ($M/L_V > $100).

The recent works of \citet{cicuendezetal2018} and \citet{cb2018} have fundamentally altered
the interpretation of Sextans’ morphology. By analyzing wide-field DECam \citep{flaugheretal2015}
photometry and Magellan/MMFS spectroscopy, these authors uncovered a ring-like overdensity 
located at a radial distance of 6\arcmin - 14\arcmin (approximately 0.1$^o$ to 0.2$^o$) from 
the galactic center. This structure is not merely a spatial density fluctuation; it is 
kinematically distinct, exhibiting a radial velocity offset of $\sim$2.5 km/s and a subtle 
Magnesium (Mg) abundance deficit ($\sim$0.03 dex) compared to the core.

Crucially, \citet{cb2018} demonstrated that appearances can be deceiving 
regarding the color-metallicity relationship in Sextans. While traditional analyses 
separate blue (metal-poor) and red (metal-rich) red giant branch stars to map different populations, 
spectroscopic follow-ups reveal that both color-magnitude diagram (CMD) regions are heavily 
contaminated by a mix of metallicities. The spatial distribution of these populations 
overlaps significantly, particularly in the 6\arcmin - 14\arcmin ring region, implying that 
the ring is a composite structural feature rather than a chemically isolated one.
The existence of such a ring, coupled with a NE shell-like overdensity and the 
galaxy’s high ellipticity, provides strong evidence for a past accretion or minor merger
 event. In a merger scenario, the shredded remains of a satellite galaxy often form 
caustics or shells at the orbital apocenters \citep{quinn1984,jonhstonetal2008}. In 
Sextans, the presence of these substructures suggests that a smaller, possibly more 
metal-poor system was swallowed by a larger, pre-existing host galaxy.

Current data leaves a critical gap in our understanding: the timing of this event and 
the specific chemical nature of the progenitor. While integrated spectra
\citep{battagliaetal2011,kirbyetal2011} as well as high-dispersion spectroscopy
of individual stars \citep{theleretal2020,roedereretal2023,tolstoyetal2025,yangetal2025}
suggest a range of metallicities, the high-resolution star formation history (SFH) and age-metallicity 
relationship (AMR) across different radial bins remain essential to determine if the
ring is a dynamical ripple of the host’s stars or the actual debris of a chemically 
distinct invader. In this paper, we utilize a full CMD-fitting approach to 
reconstruct the SFH and AMR across four radial zones, namely: core, ring, outer body, 
and outskirts, to disentangle the temporal and chemical signals of this proposed merger.
In Section~2, we describe the data used and the filtering procedure to produce a clean
Sextans's CMD. Section~3 deals with the derivation of spatially resolved SFRs, AMRs and 
cumulative mass functions, while Section~4 discusses the present results. Finally,
Section~5 summarizes the main conclusions of this work.

\begin{figure}
\includegraphics[width=\columnwidth]{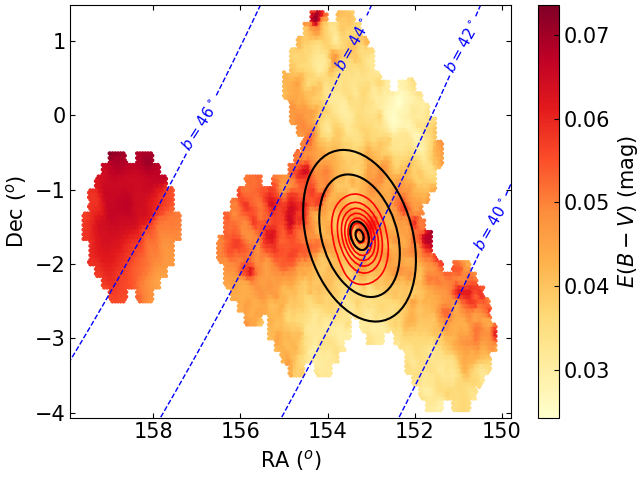}
\caption{Interstellar reddening ($E(B-V)$) map across the field of
Sextans,  centered at (RA,Dec) = (153.26$^o$,-1.61$^o$). Iso-density contours
are superimposed with red curves, while ellipses with semi-major axes of 6\arcmin, 
14\arcmin, 60\arcmin and 84\arcmin are plotted with black curves, respectively. Blue
dashed lines represent different Galactic latitudes. The single DECam field
centered at  (RA,Dec) $\sim$ (158.5$^o$,-1.1$^o$) represents the reference field
(see text for details).}
\label{fig1}
\end{figure}

\section{Data handling}

The primary photometric data for this study were retrieved from the wide-field survey 
conducted by \citet{cicuendezetal2018}. These observations were obtained 
using the Dark Energy Camera \citep[DECam;][]{flaugheretal2015}, an extremely wide-field
 570-megapixel imager mounted at the prime focus of the Victor M. Blanco 4-m telescope 
at the Cerro Tololo Inter-American Observatory (CTIO). DECam’s 3 deg$^2$ field of view 
and high sensitivity make it uniquely suited for tracing the low-surface-brightness 
outskirts of Milky Way satellites like Sextans.

The catalog \citep[Table~2;][]{cicuendezetal2018} comprises deep $g$ and $r$ photometry. 
A critical characteristic
 of this dataset is its depth and completeness; the photometry reaches a 50\% completeness 
level at approximately $g \approx$ 24.5 mag and $r \approx$ 24.1 mag. This depth is 
essential for resolving the Main Sequence Turnoff (MSTO) at the distance of Sextans 
($(m-M)_o \approx$ 19.67 mag), allowing for a
 high-fidelity reconstruction of the star formation history. Photometric errors are 
remarkably low in the regions of interest, typically $\la$ 0.02 mag at the level of the Red
 Giant Branch (RGB) and remaining below 0.1 mag even at the 50\% completeness limit.

\begin{figure*}
\includegraphics[width=\columnwidth]{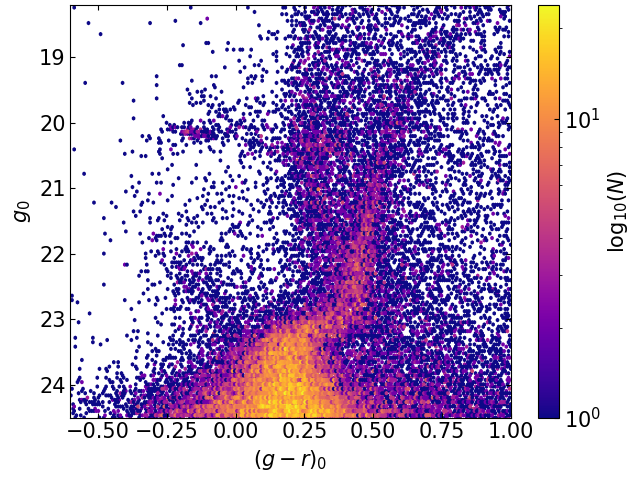}
\includegraphics[width=\columnwidth]{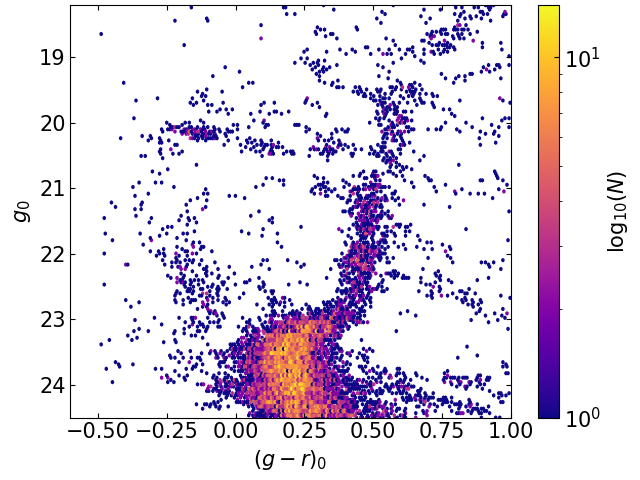}
\caption{Sextans CMDs before (left) and after (right) the decontamination of
MW field stars.}
\label{fig2}
\end{figure*}

To ensure the integrity of the stellar sample, we utilized the morphological 
classification provided in the \citet{cicuendezetal2018} catalog. DECam data reduction 
pipelines assign a  morphological type to each detected source based on its spread-model 
parameter. We adopted a stringent first filter by selecting only sources with a morphological 
type of -1, which corresponds to objects classified as point-like (stars) with high confidence.
Table 2 of \citet{cicuendezetal2018} contains a total of 602.019 detections. After 
applying the morphological type -1 filter to eliminate background galaxies and 
instrumental artifacts, we retained a sample of 281.948 stars. This selection ensures
 that the resulting CMDs are not contaminated by unresolved 
distant galaxies, which can mimic the colors of faint Main Sequence (MS) stars.

The Sextans dSphs is located at a relatively low Galactic latitude ($l$, b) = 
(243.5$^o$,+42.3$^o$), making it susceptible
to non-negligible and potentially differential interstellar extinction. To account for 
this, we utilized the individual $E(B-V)$ values provided in \citet{cicuendezetal2018} 
(Table~2), which were computed for each star according to the
\citet{sf11} maps.
Figure~\ref{fig1} illustrates the spatial distribution of the DECam fields used in this 
study, with the color-coding representing the individual $E(B-V)$ values. The presence of 
spatial gradients in the reddening map confirms the necessity of using individual
 corrections rather than a global average for the galaxy. We transformed the observed 
magnitudes to intrinsic $g_0$ and $(g-r)_0$ values using the extinction 
coefficients derived for the DECam filters.
These coefficients allow us to construct dereddened CMDs that are essential for the 
subsequent CMD fitting analysis. By correcting for differential reddening at
the individual stellar level, we minimize the artificial broadening of the MSTO and 
RGB, thereby increasing the temporal and chemical resolution of the recovered SFH.

A primary challenge in the analysis of the Sextans dSph is the severe contamination 
from Milky Way (MW) foreground stars. At the Galactic coordinates of Sextans, the line 
of sight passes through significant components of the Galactic disk and halo, resulting 
in a dense population of field stars that overlap with the faint Sextans MS. To isolate 
the galaxy’s members, we adopted a probabilistic, likelihood-based statistical subtraction 
method that treats the CMD as a continuous density surface rather than a discrete grid.
Following the data retrieval and morphological filtering described above, the data set 
was partitioned into two distinct spatial regions to facilitate background subtraction,
namely: the region centered on the Sextans dSph (RA$<$157$^o$), and a nearby control 
region used to sample the pure Milky Way (MW) foreground (RA$>$157$^o$) (see
Figure~\ref{fig1}).
Because the total sky area covered by the Sextans region is significantly larger than 
that of the reference region, a fixed scaling factor of $k$=6 was applied to normalize the 
reference star counts to the Sextans area. This normalization is strictly geometric,
corresponds to the ratio of the areas covered by the Sextans and reference fields,
and is required to perform a statistically valid comparison of the stellar densities in 
the CMD. Note that the background subtraction was done for the entire 
Sextans field (RA$<$157$^o$) at once.

For both fields, we constructed a two-dimensional histogram in the reddening corrected 
CMD. To address Poisson noise and the scarcity of stars in sparsely populated 
regions, we applied Gaussian kernel smoothing. This technique transforms discrete stellar 
counts into a continuous density surface, representing the probability density of 
finding a star at any given point in the CMD for both the Sextans ($D_{Sextans}$) and the
 reference ($D_{ref}$) populations.
For every star $i$ in the Sextans field, we calculated a membership likelihood ($P_i$). This
 probability is derived from the local density ratio:

\begin{equation}
P_i = \max\left(0, 1 - \frac{k \times D_{\text{ref}}(\text{color}_i, \text{mag}_i)}{D_{\text{Sextans}}(\text{color}_i, \text{mag}_i)}\right)
\end{equation}

\noindent where $P_i$ $\approx$ 1 indicates a high probability of membership in Sextans, and 
$P_i$ $\approx$ 0 indicates a likely foreground contaminant. This approach is intrinsically
Bayesian; we assume a prior where the MW foreground is omnipresent. By using
smoothed density maps, every bin maintains a non-zero density value, preventing 
the model from becoming overconfident in regions where the reference field might 
have zero stars due to random sampling effects.

Rather than applying a hard probability cut (e.g., $P_i$ $>$ 0.5), which can introduce
artifacts and eliminate real members in low-density regions, we performed a stochastic 
selection process. For each star, a random number was drawn from a uniform distribution 
$U[0,1]$;  the star was retained only if the random number was less than $P_i$. This 
preserves the underlying statistical distribution of the galaxy’s population, 
while effectively thinning the population according to the local contamination level.
After applying this decontamination procedure to our 245.657 filtered
 stars in the Sextans region (RA $<$ 157$\degr$), the sample was reduced to 41.656 high-confidence 
Sextans members, whose iso-density contours are superposed in Figure~\ref{fig1}.
This method significantly enhanced the contrast of the 
Sextans RGB and MS. The resulting decontaminated catalog provides a robust, 
high-fidelity foundation for the subsequent SFH) and AMR analysis. Figure~\ref{fig2}
illustrates the MW decontamination procedure,  while Table~\ref{tab1} summarizes the
performance of the decontamination method. As can be seen, because the likelihood method 
operates in color-magnitude space rather than physical space, the larger spatial crowding 
of the galaxy core forces a higher volume of overlapping stars to be stripped out per unit 
area. This is why the final calculated density of removed stars peaks sharply at the center, 
even though the physical background contamination itself is nearly constant across 
the analyzed field. Figures similar to Figure~\ref{fig2}
for the regions devised in Section~3 are shown
in the Appendix (see Figures~8-11).

The statistics provided in Table~\ref{tab1} reveal a radial 
gradient in the decontamination efficiency. By comparing the number of 
removed stars (observed minus cleaned) to the expected foreground 
contribution derived from the reference field, it is evident that the 
inner regions (core and ring) undergo more significant statistical 
thinning than the outskirts. For example, while the expected foreground 
in the core is $\sim$233 stars based on a uniform density assumption, 
the procedure removes 1.438 stars. Conversely, in the outskirts, 
the number of removed stars ($\sim$23.932) is closely aligned with the 
expected background ($\sim$25.590).

This systematic variation is a direct consequence of the likelihood-based 
subtraction method defined in Equation~(1). Because the membership 
probability $P_i$ is inversely proportional to the local Sextans density 
$D_{\rm Sextans}$, stars in high-density regions of the CMD (e.g., the core's 
crowded MSTO and RGB) are assigned higher individual probabilities. 
However, as discussed above, the larger spatial crowding in the 
center means that even a minor statistical contribution from the 
Galactic foreground results in a higher absolute volume of overlapping 
stars being stripped out to match the aggregate dwarf galaxy density 
\citep[see also][]{cicuendezetal2018}. Essentially, the method prioritizes 
suppressing the Milky Way noise floor in regions where the Sextans 
signal is strongest to avoid artificial broadening of evolutionary sequences.

To ensure this variable efficiency does not bias the recovered SFHs or 
cumulative mass assembly, we rely on two key safeguards. First, the 
validation against 3D membership probabilities from \citet{tolstoyetal2025} 
demonstrates that the routine successfully identifies the structural 
loci of the galaxy, retaining true members at three times the rate of 
non-members even in thinned regions (see below). Second, 
the \texttt{PANCAKE} CMD-fitting code utilizes a Poisson-based likelihood 
statistic \citep{d02}. This formulation (see Equation 3 below) explicitly 
accounts for the reduced stellar counts in decontaminated bins by 
incorporating correct Poisson weights, thereby preventing the lower 
sample sizes in the inner regions from introducing systematic offsets in 
the recovered mass weights. Consequently, the derived AMR bifurcations 
and assembly timescales are driven by the high-confidence signal of 
the retained population rather than being an artifact of the 
decontamination gradient.

With the aim of quantifying the performance of the decontamination procedure,
we used, as far as we are aware, the largest sample of stars with classical 3D
membership probabilities \citep{tolstoyetal2025}. To compare them with our density-driven 
color-magnitude likelihood values, we cannot perform a direct star-by-star equality match, 
because they represent entirely different concepts: the former being a physical 
confirmation probability for individual stars and ours being a local overdensity weight 
for statistical populations. Instead, we compared them using statistical cross-examination 
frameworks. \citet{tolstoyetal2025} provided two tables (D.1 and D.2) with 559 members
and 434 non-members, respectively, included in our photometric data set. We used 
the TOPCAT software \citep{taylor2005} to match tables using (RA,Dec) coordinates with
a tolerance of 1\arcsec, and found that 30\% and 10\% of the stars in D.1 and D.2 tables,
respectively, are among the cleaned CMD's stars. This outcome shows that the
decontamination procedure successfully discriminated populations. The survival rate for 
members is three times the survival rate of non-members. This proves that the CMD-density 
matrix holds significant predictive power; it successfully identifies where the structural 
features of Sextans (e.g., MSTO, RGB) are in the CMD space, preferentially keeping true stars.
The 10\% of \citet{tolstoyetal2025}'s non-members that the routine kept are Milky Way 
foreground mimics. Because the routine works strictly on color and magnitude, if a Milky 
Way halo star happens to have the exact same color and magnitude as a Sextans RGB star, 
the code cannot distinguish it. It assigns it a high probability because 
it falls into a pixel where Sextans stars are dense.
The decontamination procedure carried out a 
statistical CMD-subtraction routine designed to blindly but effectively 
thinning the catalog down to match the aggregate population density of the dwarf galaxy,
and successfully suppress non-members while favoring real members. We found that
the median of the CMD membership probability weights 
 ($P_i$, see Equation~1) resulted to be 0.44 for both
the confirmed spectroscopic members and non-members from
\citet{tolstoyetal2025}, thus reflecting the local signal-to-noise ratio of the data grid, making 
it a perfect tool for population-wide SFH fitting.

 A weight of 0.44 implies that, in these specific CMD bins, the normalized 
MW foreground contributes approximately 56\% of the total observed 
stellar density ($1 - 0.44 = 0.56$). This characterizes the challenging 
nature of the decontamination in regions like the RGB, where 
Sextans stars and MW halo stars overlap significantly in color and 
magnitude.
Second, the fact that the median weight is identical (0.44) for both 
populations is a critical validation of our statistical approach. It 
demonstrates that our membership selection is truly blind to the 
physical nature of individual stars and is driven solely by the 
morphology of the CMD density matrix. Non-members are not 
assigned lower probabilities because they are non-members; rather, they 
remain in the sample as foreground mimics because they reside in 
the same high-density CMD loci as true Sextans stars. This 
uniformity in the median weight ensures that the subsequent population-wide 
SFH fitting is not biased by preferential selection but instead 
operates on a statistically thinned catalog that accurately matches 
the aggregate density of the dwarf galaxy.

\begin{deluxetable*}{lccc}
\tablecaption{Statistics of observed and cleaned stars (see Section 3 for
details about the different devised elliptical regions).}
\label{tab1}
%\tablewidth{0pt}
\tablehead{\colhead{Region} & \colhead{Area (deg$^{\rm -2}$)} & \colhead{Observed stars} & \colhead{Cleaned stars} }
\startdata
Reference field (RA $>$ 157$\degr$) & 3.12 & 36.291 & --- \\
Sextans field (RA $<$ 157$\degr$)    & 18.02 & 245.657 & 41.656\\
0\arcmin-6\arcmin (core) &           0.02 &  2.112      &  674 \\ 
6\arcmin-14\arcmin (ring) &          0.10 &  6.992      &  2.054 \\
14\arcmin-60\arcmin (outer body) &   2.17 &  41.263     &  9.100 \\
60\arcmin-84\arcmin (outskirts) &    2.20 &   28.468    &  4.536 \\\hline
\enddata
\end{deluxetable*}

\begin{figure}
\includegraphics[width=\columnwidth]{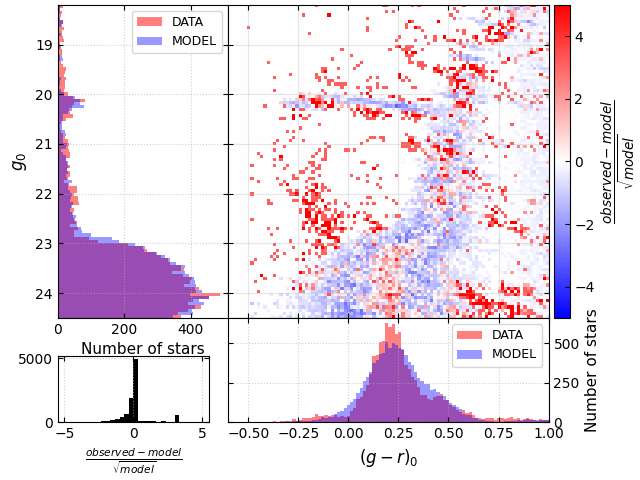}
\caption{Diagnostic CMD-fitting results for the Sextans dSph
 for one of the bootstrapping realizations. The top-left panel 
shows luminosity distribution ($g_0$) comparing observed data (red) and synthetic 
model (blue).The top-rigth panel displays the significance of the residuals, defined as 
$(observed - model) / \sqrt{model}$. The bottom panels provide the residuals 
distribution (left) and the color distribution $(g-r)_0$ (right) comparing the data 
(red) and model (blue).}
\label{fig3}
\end{figure}

\section{The Sextans SFH}

To derive the SFH and chemical enrichment properties of Sextans, we utilize the 
\texttt{PANCAKE} code \citep[Python-based Numerical Color-magnitude-diagram 
Analysis pacKagE;][]{zhengetal2025}. \texttt{PANCAKE} is an open-source framework 
designed to perform non-parametric SFH recovery by comparing observed CMDs with 
synthetic populations generated from theoretical isochrones. Our analysis relies 
on a grid of simple stellar populations (SSPs) generated using the PARSEC isochrone 
library \citep{betal12}. The grid spans a range of ages from log(age /yr) = 6.6 to 
10.15 and metallicities [Fe/H] from -2.9 dex to +0.5 dex. Each SSP is populated assuming 
a \citet{kroupa02} initial mass function and an assumed binary fraction of 
$f_b$ = 0.35. 

A significant challenge in CMD fitting is the accurate characterization 
of observational effects. In the original version of the code, error modeling was 
addressed through the reading of input artificial star test (ASTs).
In this work, we have implemented a more robust treatment of 
completeness and photometric uncertainties. Instead of relying on a sparse set of 
artificial star tests, we incorporate an externally derived error and completeness model. 
This model, taken from \citet{cicuendezetal2018}, provides the standard deviation of  
magnitudes ($\sigma_g, \sigma_r$) and the detection probability (completeness) as 
a continuous function of magnitude and color. By injecting these parameters directly 
into the template construction phase, we ensure that each synthetic SSP, is convolved 
with the exact observational footprint of the data. 

The fundamental principle of the code is to model the observed stellar distribution in 
the CMD as a linear combination of synthetic SSPs. 
Mathematically, the predicted number of stars in the $j$-th CMD bin, $M_j$, is defined as:
\begin{equation}
    M_j = \sum_{i=1}^{N} w_i T_{i,j}
\end{equation}
where $T_{i,j}$ represents the density of stars in the $j$-th bin for the $i$-th SSP template 
(of a given age and metallicity), and $w_i$ is the non-negative weight (mass) assigned to
that population. The code utilizes a linear programming solver to minimize the objective 
function, defined by a Poisson-based likelihood, to recover the star formation 
rate (SFR) as a function of lookback time and metallicity.

Unlike traditional approaches that adopt literature values for distance and extinction, 
we have modified the \texttt{PANCAKE} optimizer to treat the distance modulus $(m-M)_0$ 
as a free parameter,  spanning values from 19.50 mag to 19.90 mag in steps of 0.05 mag
(the CMDs is corrected by $E(B-V)$).  The resulting best-fitted $(m-M)_0$ turned out to
be 19.70 mag, in very good agreement with \citet[][19.67 mag]{cicuendezetal2018}. We perform a 
systematic grid search 
across the parameter space, identifying the global minimum of the Poisson likelihood 
statistic \citep{d02}:

\begin{equation}
    \chi^2_{\lambda} = 2 \sum_{j} \left[ M_j - O_j + O_j \ln\left(\frac{O_j}{M_j}\right) \right]
\end{equation}

\noindent where $O_j$ is the observed count in the $j$-th bin and $M_j$ is the model 
prediction for a given age, metallicity, and distance. This formulation 
ensures that bins with low stellar counts are treated with correct Poisson weights, 
avoiding the biases inherent in standard Gaussian least-squares approaches. By 
identifying the minimum of this function, we ensure that the subsequent SFH is not 
biased by incorrect structural assumptions. To maintain optimal structural resolution 
across the CMD, we employ an adaptive binning strategy based on centroidal Voronoi 
tessellation. We set a minimum threshold of 100 stars per bin to ensure that the 
Poisson uncertainty remains 10\% in every data point. The resulting binning nodes 
are used to map both the observed stars and the synthetic templates into a consistent 
manifold for the linear solver. Furthermore, to estimate formal 
uncertainties on the resulting SFR and AMR, we implement a bootstrap resampling 
technique. By creating multiple realizations of the observed catalog and re-fitting 
each, we derive the $16^{th}$ and $84^{th}$ percentiles of the distributions to 
represent the $1\sigma$ confidence intervals.

Given the structural complexity reported in previous studies of Sextans dSph, including 
the possibility of a stellar ring and signatures of a past accretion event 
\citep{cb2018}, we adopt a spatially resolved analysis strategy. The galaxy 
is divided into four concentric elliptical regions defined by their semi-major axis
($r$), namely: $r <$ 6\arcmin (core); 6$\arcmin$ $< r <$  14$\arcmin$ (ring);  
14$\arcmin$ $< r <$  60$\arcmin$ (outer body); and 60$\arcmin$ $< r <$  84$\arcmin$ 
(outskirts),  respectively (see Figure~\ref{fig1}). The geometry of these ellipses (position angle $\theta = 
56^\circ$ and ellipticity $\epsilon = 0.27$) is fixed according to the structural 
parameters determined by \citet{cicuendezetal2018}. By fitting each region independently, 
we can track the radial migration of star formation and search for localized variations 
in the chemical enrichment history.

\begin{figure*}
\includegraphics[width=\columnwidth]{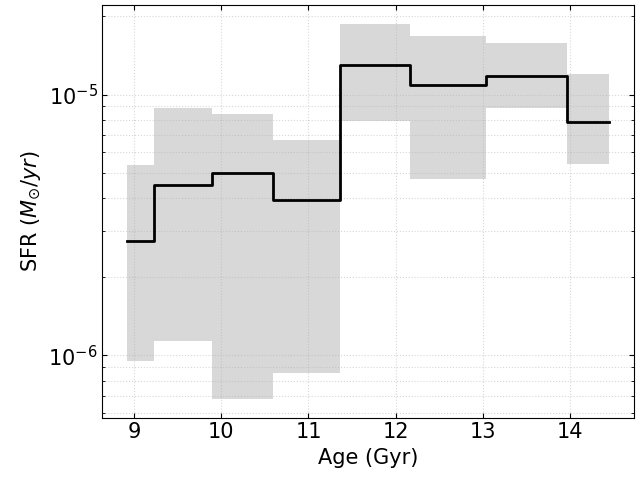}
\includegraphics[width=\columnwidth]{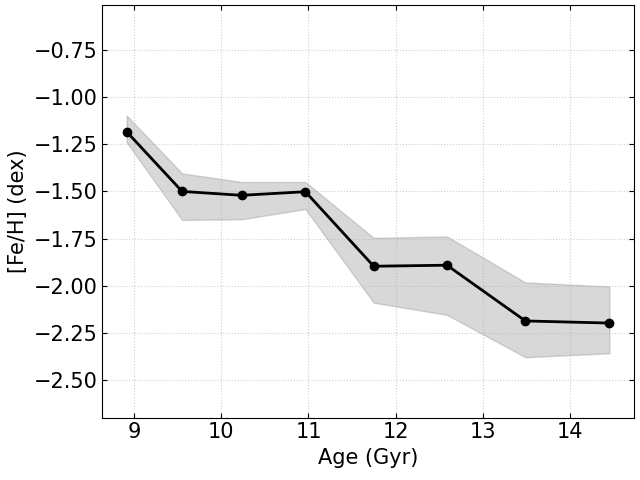}
\caption{SFR (left panel) and AMR (right panel) of the Sextans dSph. The
gray shades represent the 1$\sigma$ confidence levels.}
\label{fig4}
\end{figure*}

To evaluate the reliability of our recovered SFH, we perform a 
detailed comparison between the observed and synthetic CMDs 
(Figure~\ref{fig3}, and also similar figures for the four devised regions
(see Figures~12-13 in the Appendix)). 
The residual map (top-right panel) indicates that the \texttt{PANCAKE} model 
successfully reproduces the primary morphological features of Sextans, including the 
MS, the MSTO, and the red giant branch (RGB). 
Quantitatively, the significance of the residuals remains largely within the 
$\pm$3$\sigma$ range across the majority of the CMD manifold. The bottom-left panel 
confirms this statistical consistency, showing a residual distribution that is 
strongly peaked near zero. Furthermore, the 1D luminosity (top-left panel) and color 
distributions (bottom-right panel) demonstrate a high degree of fidelity in reproducing 
the integrated stellar density, with the synthetic model (blue) closely tracing the 
observed counts (red) across the full color range of $g_0$ and $(g-r)_0$. These 
results suggest that our treatment of photometric uncertainties and the adaptive Voronoi 
binning scheme effectively minimize systematic biases in the SFH recovery.

To properly interpret the diagnostic features shown in the two-dimensional 
residual map (Figure~\ref{fig3}, top-right panel), an essential distinction must be made 
regarding the nature of the data points: the panel maps discretized grid cells (or pixels) 
resulting from our phase-space binning configuration, rather than individual stars. 
Consequently, any given CMD cell that deviates from the model occupies an identical 
physical pixel area on the plot, irrespective of whether that specific cell contains a 
single isolated star or a high-density stellar population.
This grid configuration explains why a small number of localized, structurally aligned 
pixels, such as those following the geometry of the MSTO, the horizontal branch (HB), or 
the RGB, appear visually prominent, while the global parameters remain well-constrained. 
The marginal nature of these local variations is robustly demonstrated by the supporting 
diagnostic panels of Figure~\ref{fig3}.

The coherent pixel structures primarily highlight well-documented systematic limitations
common to all synthetic CMD techniques, typically stemming from minor discrepancies in 
stellar evolutionary tracks (e.g., slight variations in color-temperature transformations, 
helium fractions, or mixing-length calibrations).
Furthermore, a cluster of highly positive residuals is visible in the upper MS region, 
which we interpret as a population of Blue Straggler Stars (BSS) since they are 
ubiquitous in ancient stellar systems like Sextans \citep{mateoetal1991}.  This mismatch 
represents an intentional boundary condition of our modeling framework. Our synthetic 
CMD template grids are generated using standard single-star stellar evolutionary tracks 
to isolate and resolve the primary SFH and AMR of the system. Because BSS are predominantly 
exotic products of binary evolution pathways, they cannot be natively synthesized by a 
single-star grid. Additionally, because the relative residual calculation scales inversely 
with the model density, sparse regions of the CMD parameter space where the model count 
approaches zero are mathematically prone to severe pixel color saturation even when 
encountering only one or two isolated empirical stars. Because the BSS population operates 
on a decoupled evolutionary mechanism, its presence does not introduce a statistical bias 
into the core SFH or AMR solutions derived from the primary, single-star stellar populations.

The SFH and AMR presented in Figure~\ref{fig4} are calculated using an optimization grid 
explicitly restricted to old stellar  populations, with isochrone components spanning 
log(age /yr) $>$ 9.95. Younger age ranges were excluded from 
the initialization configuration as they introduce hot MS features that are 
structurally incompatible with the observed CMD and degrade the 
reproduction of the integrated 1D luminosity function. No star formation values or 
corresponding error bars exist below this threshold as they lie outside the defined parameter 
domain of the model.

\section{Analysis and discussion}

The global archaeological reconstruction of Sextans dSph, presented in 
Figure~\ref{fig4}, confirms its status as a predominantly ancient system, dominated 
by a massive star forming event between 12 and 14 Gyr ago. The SFH shows a classic 
rapid initial burst that accounts for the bulk of the galaxy's stellar mass, 
followed by a sharp decline in activity. This global profile is consistent with 
the fossil paradigm proposed by \citet{bettinellietal2018}, suggesting that 
Sextans underwent rapid early assembly before environmental quenching, 
abruptly halted further growth approximately 1.3 Gyr after the Big Bang.
The integrated AMR in Figure~\ref{fig4} (right panel) 
reveals a progressive chemical enrichment starting from [Fe/H] $\approx$ -2.5 dex. 
Interestingly, the enrichment slope remains relatively flat along the whole age range. 
This flat AMR at early times is a key 
feature in low-mass dwarf spheroidals where shallow potential wells struggle 
to retain metals during intense bursts of star formation.

While the global SFH suggests a simple early-quenching scenario, the structural 
and kinematic evidence suggests a more turbulent past. \citet{cicuendezetal2018} 
and \citet{cb2018} identified a distinct stellar ring-like overdensity at 6\arcmin - 
14\arcmin from the galactic center. Our analysis confirms that 
this structure is not merely a spatial density fluctuation but a signature of a 
complex assembly history.
The ring exhibits two critical signatures identified in previous literature 
that point toward a minor merger or accretion event: a radial velocity offset 
of $\sim$2.5 km/s and a subtle Magnesium (Mg) abundance deficit ($\sim$0.03 dex) compared 
to the core. In the context of galactic archaeology, a localized Mg deficit 
is a chemical tag indicating that the stars in the ring formed in an 
environment with lower star formation efficiency than the host core, i.e.,
a hallmark of stars born in a smaller, separate dark matter halo before being incorporated 
into the larger system.

\begin{figure}
\includegraphics[width=\columnwidth]{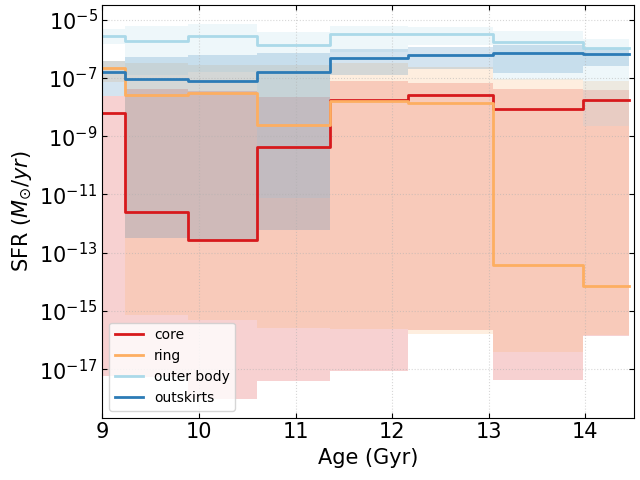}
\caption{Spatially resolved SFR as a function of age for the four designated 
elliptical regions: core (red), ring (orange), outer body (cyan), and outskirts 
(blue). Shades represent 1$\sigma$ uncertainties as extracted from the
\texttt{PANCAKE}'s metadata bootstrapping results. Note the distinct 13 Gyr onset of the ring 
population and the sustained, 
high-level SFR of the outer body, which represents the primary mass-assembly 
site of the system. The core and outskirts show early declines ($\sim$11.5 Gyr), 
while the merger-associated regions (ring and outer body) persist until 9 Gyr.}
\label{fig5}
\end{figure}

\begin{figure}
\includegraphics[width=\columnwidth]{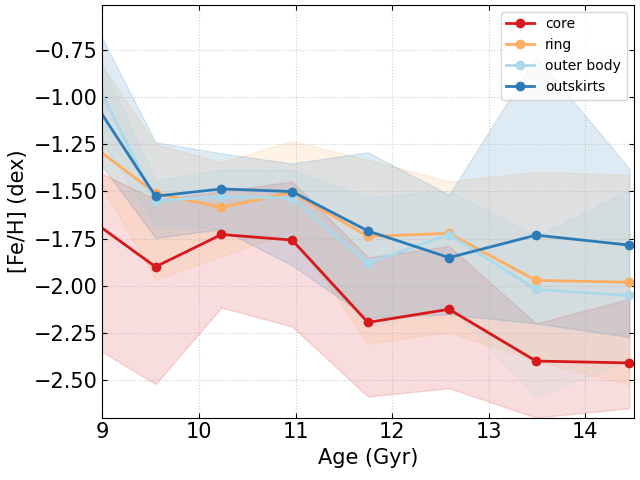}
\caption{AMR for the four radial zones. Shades represent 1$\sigma$ uncertainties as extracted from the
\texttt{PANCAKE}'s metadata bootstrapping results. The core (red) follows a smooth enrichment 
trend but remains systematically more metal-poor by $\Delta$[Fe/H] $\approx$0.5 
dex compared to the other three regions. The ring (orange), outer body (cyan), 
and outskirts (blue) show nearly identical, more metal-rich enrichment histories.
This offset indicates that the merger event introduced a significant reservoir of 
chemically evolved gas that fueled star formation throughout the majority of the 
galaxy's volume.}
\label{fig6}
\end{figure}

\begin{figure}
\includegraphics[width=\columnwidth]{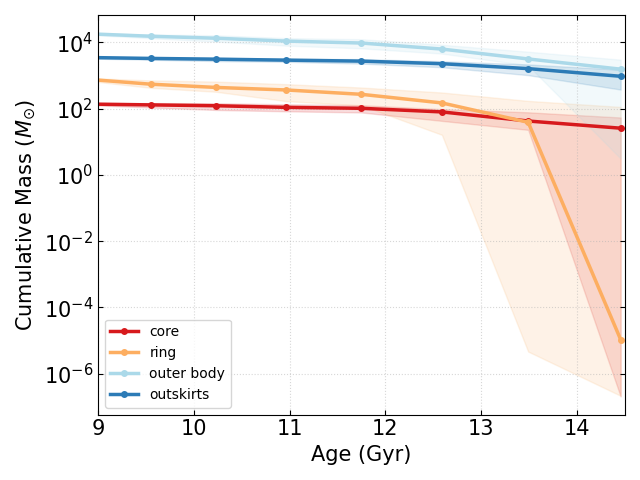}
\caption{Cumulative stellar mass assembly as a function of age for the four radial 
zones. Shades represent 1$\sigma$ uncertainties as extracted from the
\texttt{PANCAKE}'s metadata bootstrapping results.  The core (red) shows the most rapid assembly, reaching its mass plateau 
early, while the outskirts (blue) exhibit a protracted growth history, reaching 
their final mass nearly 1 Gyr later. This radial gradient in assembly time is indicative 
of inside-out growth and is consistent with quenching driven by environmental 
interactions with the Milky Way halo rather than reionization alone.}
\label{fig7}
\end{figure}

The fundamental challenge in interpreting the assembly of the Sextans dSph lies 
in determining whether its complex structural features, specifically the 
6\arcmin - 14\arcmin ring identified by \citet{cb2018}, formed in situ or were 
acquired through a late-time accretion event. By leveraging the spatially 
resolved capabilities of \texttt{PANCAKE}, we compare the SFR (Figure~\ref{fig5}), 
the AMR (Figure~\ref{fig6}), and the cumulative mass assembly (Figure~\ref{fig7}) 
across four distinct radial bins.

The spatially resolved SFR across the four elliptical regions of Sextans dSph
(Figure~\ref{fig5}) reveals a complex, non-monolithic assembly history. The 
primary differentiator is the temporal decoupling between the core (red line) and 
the ring (orange line), and the sustained activity in the outer body (cyan)
with respect to other galaxy regions.
The core shows a robust and early onset of star formation, maintaining high 
levels from approximately 14 Gyr until a sharp decline at 11.5 Gyr. This represents 
the original, receiving dwarf galaxy; a concentrated stellar seed that began forming 
stars at the earliest possible cosmic epoch. In contrast, the ring population 
exhibits a distinct delayed onset. Star formation in the ring does not begin 
until $\sim$13 Gyr, nearly 1 Gyr after the core's inception. Crucially, the ring 
maintains a steady SFR with respect to other regions until 9 Gyr, significantly 
outlasting the core's primary burst.

This 13 Gyr timestamp is a critical diagnostic. The appearance of the ring's 
SFR, occurring while the core was already active, strongly suggests an 
injection event. In the context of current $\Lambda$CDM literature 
\citep[e.g.][]{benitezllambayetal2016,amorisco2017}, such a localized, 
delayed overdensity is most likely the result of a minor merger. The receiving 
core galaxy merged with a secondary, gas-rich system whose orbital energy 
and subsequent tidal disruption deposited material specifically at the 
6\arcmin - 14\arcmin radial distance, creating the observed ring-like structure.
A key finding of this work is that the outer body (cyan), rather than the core, 
serves as the dominant reservoir of stellar mass. The outer body exhibits a 
remarkably constant and higher level of SFR from 14 Gyr down to 9 Gyr
with respect to the other regions. This 
implies that while the merger event (the ring) and the initial seed (the core) 
were localized, the galaxy as a whole was undergoing massive, distributed 
star formation. The injection of the merging satellite at 13 Gyr likely provided 
a dual effect: it created the spatial overdensity of the ring, and it fueled the 
continued star formation in the outer body and outskirts.

The outskirts (blue) show an intermediate level of SFR, which begins to 
decline at 11.5 Gyr, synchronized with the decline of the core. This suggests 
that the quenching of the system began from the outside-in and the inside-out 
simultaneously, leaving the mid-body regions (ring and outer body) as the last 
bastions of activity until 9 Gyr. The prolonged star formation in the outer regions 
(up to 9 Gyr) places Sextans in a unique category compared to other classical 
dSphs like Draco or Ursa Minor, which are purely ancient 
\citep{aparicioetal2001,carreraetal2002}. The fact that star 
formation in the outer body and ring persists for $\sim$5 Gyr (from 14 to 9 Gyr)
 challenges the idea that Sextans was purely quenched by reionization.
Instead, our results for Figure 5 suggest that Sextans was able to retain gas 
or re-acquire it via the 13 Gyr merger event. The eventual cessation of star 
formation at 9 Gyr is more consistent with environmental stripping (e.g., an 
early infall into the Milky Way halo or a close encounter with another massive 
satellite) rather than a purely internal or UV-background driven quenching. 
The higher level of SFR in the outer body indicates that the system was 
surprisingly efficient at processing gas across a wide area before the final 
quenching took hold. On average, considering Sextans dSph as a whole, it 
more importantly formed stars until $\sim$11.5 Gyr ago (see Figure~\ref{fig4}).

By synthesizing these radial trends, we propose that the original Sextans 
was a compact system (the current core). The collision with a secondary system 
at ~13 Gyr not only deposited the stars that now form the ring but also puffed up 
the galaxy, leading to the high-volume star formation in the outer body. This 
explains why the ring and core have nearly identical SFR levels, while the much 
larger outer body area shows a higher integrated rate. This multi-component 
assembly is the only way to reconcile the distinct start-times and durations 
seen in Figure 5.

The AMR presented in Figure 6 provide the chemical evidence needed to 
refine the merger scenario identified in Figure~\ref{fig5}. Rather than a simple, 
monolithic chemical enrichment, the four radial zones exhibit a clear bifurcation 
in their chemical histories, characterized by a persistent offset in metallicity
([Fe/H]) across the entire duration of the galaxy's active life.
The AMR of the core (red line) stands as the most chemically primitive component 
of the system. It initiates at a very low metallicity and, although it follows a 
smooth enrichment trend of approximately 0.6 dex over 5 Gyr, it remains 
systematically shifted toward lower metallicities by $\Delta$[Fe/H] $\approx$
-0.5 dex relative to the other regions.
This identifies the original host or receiving galaxy as a relative metal-poor 
system. Such a low metallicity at the earliest epochs suggests that the core's 
progenitor was a low-mass dwarf with a shallow potential well, likely 
experiencing significant galactic winds that limited its early metal retention. 
The core represents the fossil remnant of this original seed, which maintained 
its distinct chemical identity even after the merger began.
Figures~\ref{fig5} and \ref{fig6} results suggest that significant star 
formation in the core ceased approximately 11.5 Gyr ago, by which time the mean 
metallicity had reached around [Fe/H] $\sim$ -2.0dex. Beyond this epoch, chemical 
evolution would be expected to be driven primarily by SNe Ia and AGB stars rather 
than by massive stars, given the much reduced star formation rate. This picture 
is broadly consistent with the detailed abundance analysis of \citet{yangetal2025}.

In contrast, the ring (orange), outer body (cyan), and outskirts (blue) exhibit 
AMRs that are nearly indistinguishable from one another but are 
more metal-rich than the core. These three regions share a parallel enrichment 
slope to the core, also rising by $\sim$0.6 dex over 5 Gyr, but starting from a 
higher baseline.
This spatial distribution suggests that the colliding galaxy, which we associated
with the onset of the ring's star formation at 13 Gyr in Figure~\ref{fig5} was a 
more massive and chemically evolved system than the original Sextans core. 
The collision did not merely trigger a localized burst; it resulted in a galaxy-wide 
injection of metal-rich gas. Once this more evolved gas was incorporated into 
the potential well, it dominated the star formation in the outer body and outskirts. 
The similarity of the AMRs in the ring, outer body, and outskirts suggests that this 
enriched gas was rapidly and efficiently mixed across the larger volume of the galaxy, 
while the core remained somewhat isolated, preserving its metal-poor signature. This 
spatial configuration is consistent with minor merger dynamics, where the accreted 
companion material does not necessarily fall into the deep potential well of the host 
core. Instead, depending on the orbital impact parameter and conservation of angular 
momentum, the tidally disrupted satellite material is preferentially deposited into 
an extended, stable ring-like structure or outer envelope 
\citep[e.g.][]{quinn1984,hw1993,mapellietal2008}. In dwarf-dwarf merger scenarios, this 
kinematics leaves the primitive core intact while embedding the chemically evolved 
accreted population into the outer radial zones \citep{amorisco2017b}.

This metal-rich accretion model provides a compelling explanation for the 
observed metallicity gradients in classical dSphs. In many dwarf galaxies, the 
core is typically more metal-rich (inside-out growth). Sextans, as shown 
here, presents the inverse: a metal-poor core surrounded by a more 
metal-rich envelope.
This configuration is a definitive signature of a  merger where the secondary 
system was larger or more efficient than the primary. Following the merger 
at 13 Gyr, the global enrichment of Sextans was essentially reset by the 
chemistry of the accreted gas. The nearly constant $\Delta$[Fe/H] offset 
between the core and the rest of the galaxy suggests that the two gas 
reservoirs (the original metal-poor gas and the newly acquired metal-rich gas) 
did not fully homogenize in the center, perhaps due to the rapid conversion 
of gas into stars or the stabilization of the core's original gas through angular
momentum conservation.

By combining the timing from Figure~\ref{fig5} (merger onset at 13 Gyr) with 
the chemical offsets from Figure~\ref{fig6}, a clear picture of the Sextans 
assembly emerges. The original Sextans was a compact, metal-poor dwarf. 
Its collision with a more massive, metal-rich satellite at 13 Gyr provided 
a fresh supply of gas and stars. This interaction puffed up the system, creating 
the outer body and the structural ring, and fueled a sustained period of star 
formation that was 0.5 dex more metal-rich than the original host. This 
dual-component nature is preserved in the chemical and temporal fossil 
records of the galaxy today.

While the SFR and AMR provide snapshots of the galaxy's activity and 
chemistry, the cumulative stellar mass assembly presented in Figure~\ref{fig7} 
quantifies the integrated growth efficiency across the different radial zones. 
This analysis is crucial for distinguishing between internal feedback 
mechanisms and external environmental influence as the primary driver of 
the galaxy's cessation of star formation.

The assembly curve for the core (red solid line) reveals a more rapid 
growth profile than the ring and the outer body, reinforcing the hypothesis of a 
compact, high-density seed that dominated the early life of the galaxy. 
Indeed, from the \texttt{PANCAKE}'s metadata bootstrapping results, 
we extracted $t_{50}$ (half-mass age) = 12.95$^{-0.78}_{+0.92}$ Gyr, 
10.82$^{-0.86}_{+1.16}$ Gyr, and 11.87$^{-1.01}_{+0.56}$ Gyr,
for the core, ring and outer body, respectively. This 
rapid assembly is consistent with the standard monolithic-like collapse of the 
innermost regions of dark matter halos, where gas densities are highest and 
star formation efficiency is maximized. By approximately 12 Gyr ago, the 
core, ring and outer region had already assembled 74.4$^{-15.5}_{+18.6}$\%, 
33.0$^{-14.8}_{+21.9}$\% and 52.5$^{-15.7}_{+14.2}$\% their stars, coinciding with 
the sharp decline in SFR seen in Figure~\ref{fig5}.

In stark contrast, the outskirts (blue line) would seem to exhibit a more lazy or 
protracted assembly history.  
The relative variation ($1\sigma/mean$) of its growth 
rate is 0.32\%, suggesting that over a 4 Gyr epoch the star formation 
and mass accumulation density were nearly flat. This scenario indicates 
that the outskirts would have reached their final mass roughly 1
 to 1.5 Gyr after the core has already quenched. This apparent temporal
 lag is a classic signature of inside-out growth, a phenomenon widely 
 observed in both simulations and observations of dwarf galaxies.
The ring (orange) and outer body (cyan) assembly curves fall between the 
core and the outskirts. Interestingly, the assembly of the outer body remains 
remarkably steady, reflecting the constant, high-level SFR identified in 
Figure~\ref{fig5}. The cumulative mass function of the ring would seem to show a subtle 
inflection around 13 Gyr, marking the point where the accretion event began 
to contribute more importantly to the local mass density. Indeed, at
13.5 Gyr its growth rate was 4.01 $\msun$/Gyr, while at 12.6 Gyr it increased
up to 12.3 $\msun$/Gyr. The fact that these 
merger-impacted regions took longer to reach their mass plateaus  would suggest
that the injected gas from the 13 Gyr collision provided the necessary 
reservoir to sustain star formation even as the central core was beginning 
to starve.

The total quenching of the galaxy by 9 Gyr across all radial bins (as seen 
by the flattening of all curves in Figure~\ref{fig7}) suggests a powerful, 
galaxy-wide termination event. In the literature, the quenching of classical 
dSphs is often attributed to cosmic reionization. However, the assembly 
curves in Figure~\ref{fig7} demonstrate that star formation persisted 
for nearly 4 Gyr after the end of the reionization epoch (z$\approx$6).
Our results are more aligned with the findings of \citet{bettinellietal2018}
and \citet{cicuendezetal2018}, which suggest that while reionization may
 have suppressed early growth, it was likely environmental stripping caused 
 by the first infall of Sextans into the Milky Way's hot gaseous corona, 
 that finally removed the remaining gas. The lazy assembly of the outskirts 
 is particularly sensitive to such external pressure; as the galaxy moved 
 through the circumgalactic medium, the loosely bound gas in the outskirts 
 would have been the first to be removed via ram-pressure stripping, 
 followed by a total cessation of activity once the supply to the inner 
 regions was cut off.

The synthesis of Figure~\ref{fig7} with the previous results suggests a
two-stage assembly process: a rapid collapse and metal-poor star formation 
in the core starting 14 Gyr ago, and the collision at 13 Gyr which puffed up 
the galaxy and added a metal-rich stellar and gas component, as evidenced 
by the sustained growth of the outer body and ring until 9 Gyr.
The divergence in the assembly times ($\sim$1 Gyr difference between core 
and outskirts) is the definitive fossil record of this transition from a compact 
seed to a more extended, merger-complex system.

Finally, we note that \citet{cb2018} found that stars located in the ring
exhibit a Magnesium (Mg) deficit of approximately 0.03 dex with respect
to the core. We utilize full CMD-fitting with the \texttt{PANCAKE} 
code, which reconstructs the SFH and AMR by analyzing the entire stellar 
population down to the MSTO. This provides a global, mass-weighted view 
of the total metallicity ([Fe/H]) based on theoretical isochrones.
In contrast, \citet{cb2018} rely on spectroscopic measurements of the Mg triplet 
($\Sigma$Mg) as a proxy for metallicity. This method is limited to a smaller 
sample (440 stars) of bright RGB and horizontal branch stars. Spectroscopy 
provides a direct measurement of specific chemical abundances, while 
CMD fitting provides a comprehensive evolutionary history of the system's 
global metallicity baseline.

It is possible for a stellar population to be relatively metal-rich in [Fe/H] 
while showing a deficit in $\alpha$-elements like Mg.  In galactic archaeology, 
an Mg deficit is often a chemical tag for stars born in environments with lower 
star formation efficiency, where the enrichment from Type Ia supernovae 
(producing iron) begins to dominate over Type II supernovae (producing Mg). 
Our finding that the ring stars are 0.5 dex more metal-rich in [Fe/H] suggests 
that the invader system was more chemically evolved overall, even if its star 
formation efficiency (reflected in the [Mg/Fe] ratio) differed from the host core.
The discrepancy may also arise from how rest of the galaxy is defined:
We isolate the core ($r<$ 6\arcmin) as a distinct region and find it to be the 
most primitive component.\citet{cb2018} used a composite control sample 
consisting of stars from both the inner and outer regions relative to the ring
($r<$ 6\arcmin and $r>$ 14\arcmin).
By specifically isolating the core, our analysis reveals a signal that may 
have been diluted in their composite control group. If the outer body and
outskirts are also metal-rich (see Figure~\ref{fig6}), comparing the ring to 
a mix of the metal-poor core and the metal-rich outskirts would naturally 
yield a different result than a direct core-to-ring comparison.

\citet{cb2018} explicitly mention that appearances can be deceiving because
 both the blue (metal-poor) and red (metal-rich) CMD regions are heavily 
 contaminated by a mix of metallicities. Their spectroscopic sample might 
 be biased toward a certain sub-population of the RGB. Our CMD-fitting 
 approach, which treats the CMD as a continuous density surface and 
 accounts for the full distribution of stars, likely provides a more representative 
 measure of the total stellar mass in each radial bin. Therefore, while 
 spectroscopy detects a subtle kinematic and $\alpha$-element signature 
 in the ring, our depth-intensive CMD fitting reveals the underlying global 
 chemical bifurcation. The inverse metallicity gradient we found: a metal-poor 
 core surrounded by a metal-rich envelope is a definitive signature of a 
 merger where a more massive and chemically evolved satellite (the ring/outer 
 body precursor) was accreted by a more primitive, low-mass host (the current 
 core). The Mg deficit in the ring found by \citet{cb2018} may simply reflect 
 the specific chemical footprint of the secondary system's own star formation 
 history before the merger.

The complex assembly history inferred from our spatially resolved SFH finds 
significant support in recent spectroscopic surveys. \citet{theleretal2020} 
and \citet{yangetal2025} both identify a distinct knee in the 
$[\alpha/Fe]$ vs. $[Fe/H]$ distribution at $[Fe/H] \sim -2.0$ dex. This 
chemical signature corresponds closely with our finding that star formation in 
the primitive core significantly declined approximately 11.5 Gyr ago, by which 
time the mean metallicity had reached this threshold. The transition 
from core-collapse supernovae dominance to Type Ia supernovae enrichment, 
noted by \citet{theleretal2020}, aligns with the secondary, more metal-rich 
star formation period we observe in the ring and outer body fueled by the 13 Gyr 
merger event.

Furthermore, the dual-component nature of Sextans is confirmed by recent 
chemo-kinematic modeling. Both \citet{tolstoyetal2025} and \citet{yangetal2025b} 
identify two distinct stellar populations: a metal-poor
 component and a metal-rich  component. While these studies generally 
find the metal-rich population to be more centrally concentrated, 
our high-resolution reconstruction would seem to suggest an inverse metallicity gradient, 
where the most primitive 
stars reside in the current core, surrounded by a chemically evolved envelope 
introduced by the accreted satellite. This discrepancy may arise from 
the specific radial definitions of the core and ring vs. the statistical 
divisions used in Jeans Anisotropic Multi-Gaussian Expansion modeling. 

Interestingly, \citet{yangetal2025} identified inflection points 
in the evolution of massive stars at $[Fe/H] \sim -2.8$ and $-2.0$. Their 
suggestion of accretion episodes and galactic winds as drivers of these shifts 
strongly corroborates our proposed 13 Gyr merger and subsequent environmental
 stripping. The puffed-up nature of the system following this merger is also 
reflected in the low inner dark matter density slope 
$(\gamma \approx 0.26)$ found by \citet{yangetal2025b}, supporting 
a core-like distribution that could be the result of the turbulent assembly 
history and subsequent stellar feedback. These synergistic findings 
between photometry and spectroscopy establish Sextans as a primary laboratory 
for studying non-monolithic dwarf galaxy evolution.

Finally, we note that Figures~8, 9 and 10 reveal a clear change in HB 
morphology, where the core is dominated by a red HB while the outer body 
exhibits a prominent blue HB. This radial trend, first characterized by 
\citet{harbecketal2011}, is qualitatively consistent with a scenario where the outer regions are 
more metal-poor than the center, seemingly aligning with traditional spectroscopic 
interpretations \citep{battagliaetal2011} and contrasting with the inverse [Fe/H] gradient 
recovered in this work.
However, HB morphology is a complex diagnostic determined by several factors beyond 
[Fe/H], including age, helium abundance, and $\alpha$-element enhancement.
For instance, in the context of stellar evolution, $\alpha$-element 
enrichment (e.g., [Mg/Fe]) significantly influences the total metallicity $Z$. A population 
that is more metal-rich in iron ([Fe/H]) but depleted in $\alpha$-elements can have a 
lower effective $Z$ than a more [Fe/H]-poor but $\alpha$-rich population. As shown by 
\citet{salarisetal1993}, $\alpha$-enhancement mimics the effect of a higher overall 
metallicity, pushing stars toward the red HB. Consequently, the primitive core, which we 
identify as an ancient ($\sim$14 Gyr), $\alpha$-rich seed, tends toward the red HB despite its 
lower [Fe/H] baseline. Conversely, the accreted satellite material in the outer regions, 
being $\alpha$-poor due to lower star formation efficiency, naturally populates the blue HB.

Although the apparent discrepancy can be reconciled within our proposed merger 
framework by considering the distinct chemical signatures of the host and the accreted satellite,
it is important to clarify that the metallicities derived from our CMD 
fitting are not direct measurements of iron abundance, but model-dependent 
quantities. They depend on the adopted PARSEC metallicity scaling, which relates 
[Fe/H] to the overall metal content $Z$ under specific abundance pattern assumptions. 
Thus, the perceived difference between CMD-derived metallicities and spectroscopic 
Mg measurements cannot be interpreted directly as [$\alpha$/Fe] without further 
validation. Furthermore, while we interpret the blue HB as tracing an ancient 
component of the accreted system, the absence of a prominent redder HB component 
from the younger, metal-rich populations inferred in the outskirts (9--12 Gyr) 
remains a point for further quantitative assessment. Until higher-resolution 
spectroscopy confirms the predicted strong radial gradient in [$\alpha$/Fe], our 
interpretation should be considered a plausible scenario rather than a definitive 
explanation.

\section{Conclusions}

Our spatially resolved archaeological reconstruction of the Sextans dSph, 
performed using deep DECam photometry and the \texttt{PANCAKE} CMD-fitting 
framework, provides a high-fidelity timeline of the galaxy's assembly. 
We summarize the primary results and proposed evolutionary scenario below.

The analysis of the SFH and AMR across four radial zones (core, ring, outer 
body, and outskirts) yields the following quantitative results:\\

$\bullet$ Sextans is a predominantly ancient 
system, with the bulk of its stellar mass formed between 12 and 14 Gyr ago. 
Star formation ceased entirely across all radial bins by 
approximately 9 Gyr ago.\\

$\bullet$ We identify a  radial 
gradient in assembly times. The core reached its mass plateau first (half-mass 
age $t_{50} = 12.95_{-0.78}^{+0.92}$ Gyr), while the outskirts exhibited a 
protracted history, reaching their mass plateau roughly 1--1.5 Gyr after the 
core. By 12 Gyr ago, the core had already assembled
 $74.4_{-15.5}^{+18.6}$\% of its stars, compared to only $33.0_{-14.8}^{+21.9}$\% 
in the ring and $52.5_{-15.7}^{+14.2}$\% in the outer  body.\\

$\bullet$ We find a inverse metallicity 
gradient. The core is the most chemically primitive component ($[Fe/H] \approx
 -2.5$ dex at inception), remaining systematically more metal-poor by 
$\Delta[Fe/H] \approx -0.5$ dex compared to the ring, outer body, and outskirts
 throughout the galaxy's active life.\\
 
 $\bullet$ The apparent discrepancy between the inverse metallicity gradient 
 and the HB morphology is addressed through a plausible scenario considering 
 $\alpha$-element variations. We propose that the ancient, $\alpha$-rich core 
 populates the red HB, while $\alpha$-poor accreted material populates the blue HB.
 
$\bullet$ Our statistical decontamination procedure was validated against spectroscopic 
catalogs, showing a membership survival rate for true members three times higher than 
for non-members. The identical median membership weights ($P_i \approx 0.44$) for both 
groups confirm that the selection is driven by the robust CMD-density matrix rather than 
preferential bias.\\
    
$\bullet$ Residuals in the upper main sequence are identified as a Blue Straggler 
population. These binary-evolution products are successfully 
isolated as intentional boundary conditions, ensuring they do not bias the recovered 
13 Gyr merger signal.\\

Based on these results, we propose that Sextans is a composite system 
formed through a minor merger. While this scenario explains several disparate 
observations, we emphasize that it relies on the assumption that CMD-derived 
metallicities accurately reflect the underlying abundance patterns. This scenario 
makes a clear, testable prediction: a strong radial gradient in [$\alpha$/Fe] 
between the central and outer regions. Future wide-field spectroscopic surveys 
will be essential to confirm these chemical signatures and validate this 
non-monolithic assembly model.

\begin{acknowledgements}
We thank the referee for the thorough reading of the manuscript and
timely suggestions to improve it.

Data used in this work are available upon request to the author.

\end{acknowledgements}

%\bibliographystyle{aasjournal}
%\bibliography{paper} % if your bibtex file is called paper.bib

\begin{thebibliography}{}
\expandafter\ifx\csname natexlab\endcsname\relax\def\natexlab#1{#1}\fi
\providecommand{\url}[1]{\href{#1}{#1}}
\providecommand{\dodoi}[1]{doi:~\href{http://doi.org/#1}{\nolinkurl{#1}}}
\providecommand{\doeprint}[1]{\href{http://ascl.net/#1}{\nolinkurl{http://ascl.net/#1}}}
\providecommand{\doarXiv}[1]{\href{https://arxiv.org/abs/#1}{\nolinkurl{https://arxiv.org/abs/#1}}}

\bibitem[{N.~C. {Amorisco}(2017{\natexlab{a}}){Amorisco}}]{amorisco2017}
{Amorisco}, N.~C. 2017{\natexlab{a}}, \bibinfo{title}{{The accreted stellar
  halo as a window on halo assembly in L$^{*}$ galaxies},} \mnras, 469, L48,
  \dodoi{10.1093/mnrasl/slx044}

\bibitem[{N.~C. {Amorisco}(2017{\natexlab{b}}){Amorisco}}]{amorisco2017b}
{Amorisco}, N.~C. 2017{\natexlab{b}}, \bibinfo{title}{{Contributions to the
  accreted stellar halo: an atlas of stellar deposition},} \mnras, 464, 2882,
  \dodoi{10.1093/mnras/stw2229}

\bibitem[{A. {Aparicio} {et~al.}(2001){Aparicio}, {Carrera}, \&
  {Mart{\'\i}nez-Delgado}}]{aparicioetal2001}
{Aparicio}, A., {Carrera}, R., \& {Mart{\'\i}nez-Delgado}, D. 2001,
  \bibinfo{title}{{The Star Formation History and Morphological Evolution of
  the Draco Dwarf Spheroidal Galaxy},} \aj, 122, 2524, \dodoi{10.1086/323535}

\bibitem[{G. {Battaglia} {et~al.}(2011){Battaglia}, {Tolstoy}, {Helmi},
  {Irwin}, {Parisi}, {Hill}, \& {Jablonka}}]{battagliaetal2011}
{Battaglia}, G., {Tolstoy}, E., {Helmi}, A., {et~al.} 2011,
  \bibinfo{title}{{Study of the Sextans dwarf spheroidal galaxy from the DART
  Ca II triplet survey},} \mnras, 411, 1013,
  \dodoi{10.1111/j.1365-2966.2010.17745.x}

\bibitem[{A. {Ben{\'\i}tez-Llambay} {et~al.}(2016){Ben{\'\i}tez-Llambay},
  {Navarro}, {Abadi}, {Gottl{\"o}ber}, {Yepes}, {Hoffman}, \&
  {Steinmetz}}]{benitezllambayetal2016}
{Ben{\'\i}tez-Llambay}, A., {Navarro}, J.~F., {Abadi}, M.~G., {et~al.} 2016,
  \bibinfo{title}{{Mergers and the outside-in formation of dwarf spheroidals},}
  \mnras, 456, 1185, \dodoi{10.1093/mnras/stv2722}

\bibitem[{M. {Bettinelli} {et~al.}(2018){Bettinelli}, {Hidalgo}, {Cassisi},
  {Aparicio}, \& {Piotto}}]{bettinellietal2018}
{Bettinelli}, M., {Hidalgo}, S.~L., {Cassisi}, S., {Aparicio}, A., \& {Piotto},
  G. 2018, \bibinfo{title}{{The star formation history of the Sextans dwarf
  spheroidal galaxy: a true fossil of the pre-reionization era},} \mnras, 476,
  71, \dodoi{10.1093/mnras/sty226}

\bibitem[{A. {Bressan} {et~al.}(2012){Bressan}, {Marigo}, {Girardi},
  {Salasnich}, {Dal Cero}, {Rubele}, \& {Nanni}}]{betal12}
{Bressan}, A., {Marigo}, P., {Girardi}, L., {et~al.} 2012,
  \bibinfo{title}{{PARSEC: stellar tracks and isochrones with the PAdova and
  TRieste Stellar Evolution Code},} \mnras, 427, 127,
  \dodoi{10.1111/j.1365-2966.2012.21948.x}

\bibitem[{R. {Carrera} {et~al.}(2002){Carrera}, {Aparicio},
  {Mart{\'\i}nez-Delgado}, \& {Alonso-Garc{\'\i}a}}]{carreraetal2002}
{Carrera}, R., {Aparicio}, A., {Mart{\'\i}nez-Delgado}, D., \&
  {Alonso-Garc{\'\i}a}, J. 2002, \bibinfo{title}{{The Star Formation History
  and Spatial Distribution of Stellar Populations in the Ursa Minor Dwarf
  Spheroidal Galaxy},} \aj, 123, 3199, \dodoi{10.1086/340702}

\bibitem[{L. {Cicu{\'e}ndez} \& G. {Battaglia}(2018){Cicu{\'e}ndez} \&
  {Battaglia}}]{cb2018}
{Cicu{\'e}ndez}, L., \& {Battaglia}, G. 2018, \bibinfo{title}{{Appearances can
  be deceiving: clear signs of accretion in the seemingly ordinary Sextans
  dSph},} \mnras, 480, 251, \dodoi{10.1093/mnras/sty1748}

\bibitem[{L. {Cicu{\'e}ndez} {et~al.}(2018){Cicu{\'e}ndez}, {Battaglia},
  {Irwin}, {Bermejo-Climent}, {McMonigal}, {Bate}, {Lewis}, {Conn}, {de Boer},
  {Gallart}, {Guglielmo}, {Ibata}, {McConnachie}, {Tolstoy}, \&
  {Fernando}}]{cicuendezetal2018}
{Cicu{\'e}ndez}, L., {Battaglia}, G., {Irwin}, M., {et~al.} 2018,
  \bibinfo{title}{{Tracing the stellar component of low surface brightness
  Milky Way dwarf galaxies to their outskirts. I. Sextans},} \aap, 609, A53,
  \dodoi{10.1051/0004-6361/201731450}

\bibitem[{A.~E. {Dolphin}(2002){Dolphin}}]{d02}
{Dolphin}, A.~E. 2002, \bibinfo{title}{{Numerical methods of star formation
  history measurement and applications to seven dwarf spheroidals},} \mnras,
  332, 91, \dodoi{10.1046/j.1365-8711.2002.05271.x}

\bibitem[{B. {Flaugher} {et~al.}(2015){Flaugher}, {Diehl}, {Honscheid},
  {Abbott}, {Alvarez}, {Angstadt}, {Annis}, {Antonik}, {Ballester}, {Beaufore},
  {Bernstein}, {Bernstein}, {Bigelow}, {Bonati}, {Boprie}, {Brooks},
  {Buckley-Geer}, {Campa}, {Cardiel-Sas}, {Castander}, {Castilla}, {Cease},
  {Cela-Ruiz}, {Chappa}, {Chi}, {Cooper}, {da Costa}, {Dede}, {Derylo},
  {DePoy}, {de Vicente}, {Doel}, {Drlica-Wagner}, {Eiting}, {Elliott}, {Emes},
  {Estrada}, {Fausti Neto}, {Finley}, {Flores}, {Frieman}, {Gerdes},
  {Gladders}, {Gregory}, {Gutierrez}, {Hao}, {Holland}, {Holm}, {Huffman},
  {Jackson}, {James}, {Jonas}, {Karcher}, {Karliner}, {Kent}, {Kessler},
  {Kozlovsky}, {Kron}, {Kubik}, {Kuehn}, {Kuhlmann}, {Kuk}, {Lahav}, {Lathrop},
  {Lee}, {Levi}, {Lewis}, {Li}, {Mandrichenko}, {Marshall}, {Martinez},
  {Merritt}, {Miquel}, {Mu{\~n}oz}, {Neilsen}, {Nichol}, {Nord}, {Ogando},
  {Olsen}, {Palaio}, {Patton}, {Peoples}, {Plazas}, {Rauch}, {Reil}, {Rheault},
  {Roe}, {Rogers}, {Roodman}, {Sanchez}, {Scarpine}, {Schindler}, {Schmidt},
  {Schmitt}, {Schubnell}, {Schultz}, {Schurter}, {Scott}, {Serrano}, {Shaw},
  {Smith}, {Soares-Santos}, {Stefanik}, {Stuermer}, {Suchyta}, {Sypniewski},
  {Tarle}, {Thaler}, {Tighe}, {Tran}, {Tucker}, {Walker}, {Wang}, {Watson},
  {Weaverdyck}, {Wester}, {Woods}, {Yanny}, \& {DES
  Collaboration}}]{flaugheretal2015}
{Flaugher}, B., {Diehl}, H.~T., {Honscheid}, K., {et~al.} 2015,
  \bibinfo{title}{{The Dark Energy Camera},} \aj, 150, 150,
  \dodoi{10.1088/0004-6256/150/5/150}

\bibitem[{D. {Harbeck} {et~al.}(2001){Harbeck}, {Grebel}, {Holtzman},
  {Guhathakurta}, {Brandner}, {Geisler}, {Sarajedini}, {Dolphin},
  {Hurley-Keller}, \& {Mateo}}]{harbecketal2011}
{Harbeck}, D., {Grebel}, E.~K., {Holtzman}, J., {et~al.} 2001,
  \bibinfo{title}{{Population Gradients in Local Group Dwarf Spheroidal
  Galaxies},} \aj, 122, 3092, \dodoi{10.1086/324232}

\bibitem[{L. {Hernquist} \& M.~L. {Weil}(1993){Hernquist} \& {Weil}}]{hw1993}
{Hernquist}, L., \& {Weil}, M.~L. 1993, \bibinfo{title}{{Spokes in ring
  galaxies.},} \mnras, 261, 804, \dodoi{10.1093/mnras/261.4.804}

\bibitem[{M. {Irwin} \& D. {Hatzidimitriou}(1995){Irwin} \&
  {Hatzidimitriou}}]{ih1995}
{Irwin}, M., \& {Hatzidimitriou}, D. 1995, \bibinfo{title}{{Structural
  parameters for the Galactic dwarf spheroidals},} \mnras, 277, 1354,
  \dodoi{10.1093/mnras/277.4.1354}

\bibitem[{M.~J. {Irwin} {et~al.}(1990){Irwin}, {Bunclark}, {Bridgeland}, \&
  {McMahon}}]{irwinetal1990}
{Irwin}, M.~J., {Bunclark}, P.~S., {Bridgeland}, M.~T., \& {McMahon}, R.~G.
  1990, \bibinfo{title}{{A new satellite galaxy of the Milky Way in the
  constellation of Sextans.},} \mnras, 244, 16P

\bibitem[{K.~V. {Johnston} {et~al.}(2008){Johnston}, {Bullock}, {Sharma},
  {Font}, {Robertson}, \& {Leitner}}]{jonhstonetal2008}
{Johnston}, K.~V., {Bullock}, J.~S., {Sharma}, S., {et~al.} 2008,
  \bibinfo{title}{{Tracing Galaxy Formation with Stellar Halos. II. Relating
  Substructure in Phase and Abundance Space to Accretion Histories},} \apj,
  689, 936, \dodoi{10.1086/592228}

\bibitem[{I.~D. {Karachentsev} {et~al.}(2004){Karachentsev}, {Karachentseva},
  {Huchtmeier}, \& {Makarov}}]{karachentsevetal2004}
{Karachentsev}, I.~D., {Karachentseva}, V.~E., {Huchtmeier}, W.~K., \&
  {Makarov}, D.~I. 2004, \bibinfo{title}{{A Catalog of Neighboring Galaxies},}
  \aj, 127, 2031, \dodoi{10.1086/382905}

\bibitem[{I. {King}(1962){King}}]{king62}
{King}, I. 1962, \bibinfo{title}{{The structure of star clusters. I. an
  empirical density law},} \aj, 67, 471, \dodoi{10.1086/108756}

\bibitem[{E.~N. {Kirby} {et~al.}(2011){Kirby}, {Lanfranchi}, {Simon}, {Cohen},
  \& {Guhathakurta}}]{kirbyetal2011}
{Kirby}, E.~N., {Lanfranchi}, G.~A., {Simon}, J.~D., {Cohen}, J.~G., \&
  {Guhathakurta}, P. 2011, \bibinfo{title}{{Multi-element Abundance
  Measurements from Medium-resolution Spectra. III. Metallicity Distributions
  of Milky Way Dwarf Satellite Galaxies},} \apj, 727, 78,
  \dodoi{10.1088/0004-637X/727/2/78}

\bibitem[{P. Kroupa(2002)Kroupa}]{kroupa02}
Kroupa, P. 2002, \bibinfo{title}{{The Initial Mass Function of Stars: Evidence
  for Uniformity in Variable Systems},} Science, 295, 82,
  \dodoi{10.1126/science.1067524}

\bibitem[{M. {Mapelli} {et~al.}(2008){Mapelli}, {Moore}, {Ripamonti}, {Mayer},
  {Colpi}, \& {Giordano}}]{mapellietal2008}
{Mapelli}, M., {Moore}, B., {Ripamonti}, E., {et~al.} 2008,
  \bibinfo{title}{{Are ring galaxies the ancestors of giant low surface
  brightness galaxies?},} \mnras, 383, 1223,
  \dodoi{10.1111/j.1365-2966.2007.12650.x}

\bibitem[{M. {Mateo} {et~al.}(1991){Mateo}, {Nemec}, {Irwin}, \&
  {McMahon}}]{mateoetal1991}
{Mateo}, M., {Nemec}, J., {Irwin}, M., \& {McMahon}, R. 1991,
  \bibinfo{title}{{Deep CCD Photometry of the Sextans Dwarf Spheroidal
  Galaxy},} \aj, 101, 892, \dodoi{10.1086/115734}

\bibitem[{M.~L. {Mateo}(1998){Mateo}}]{mateo1998}
{Mateo}, M.~L. 1998, \bibinfo{title}{{Dwarf Galaxies of the Local Group},}
  \araa, 36, 435, \dodoi{10.1146/annurev.astro.36.1.435}

\bibitem[{P.~J. {Quinn}(1984){Quinn}}]{quinn1984}
{Quinn}, P.~J. 1984, \bibinfo{title}{{On the formation and dynamics of shells
  around elliptical galaxies.},} \apj, 279, 596, \dodoi{10.1086/161924}

\bibitem[{I.~U. {Roederer} {et~al.}(2023){Roederer}, {Pace}, {Placco},
  {Caldwell}, {Koposov}, {Mateo}, {Olszewski}, \& {Walker}}]{roedereretal2023}
{Roederer}, I.~U., {Pace}, A.~B., {Placco}, V.~M., {et~al.} 2023,
  \bibinfo{title}{{Abundance Analysis of Stars at Large Radius in the Sextans
  Dwarf Spheroidal Galaxy},} \apj, 954, 55, \dodoi{10.3847/1538-4357/ace3c1}

\bibitem[{M. {Salaris} {et~al.}(1993){Salaris}, {Chieffi}, \&
  {Straniero}}]{salarisetal1993}
{Salaris}, M., {Chieffi}, A., \& {Straniero}, O. 1993, \bibinfo{title}{{The
  alpha -enhanced Isochrones and Their Impact on the FITS to the Galactic
  Globular Cluster System},} \apj, 414, 580, \dodoi{10.1086/173105}

\bibitem[{E.~F. {Schlafly} \& D.~P. {Finkbeiner}(2011){Schlafly} \&
  {Finkbeiner}}]{sf11}
{Schlafly}, E.~F., \& {Finkbeiner}, D.~P. 2011, \bibinfo{title}{{Measuring
  Reddening with Sloan Digital Sky Survey Stellar Spectra and Recalibrating
  SFD},} \apj, 737, 103, \dodoi{10.1088/0004-637X/737/2/103}

\bibitem[{J.~L. {Sersic}(1968){Sersic}}]{sersic1968}
{Sersic}, J.~L. 1968, {Atlas de Galaxias Australes}

\bibitem[{M.~B. {Taylor}(2005){Taylor}}]{taylor2005}
{Taylor}, M.~B. 2005, in Astronomical Society of the Pacific Conference Series,
  Vol. 347, Astronomical Data Analysis Software and Systems XIV, ed.
  P.~{Shopbell}, M.~{Britton}, \& R.~{Ebert}, 29

\bibitem[{R. {Theler} {et~al.}(2020){Theler}, {Jablonka}, {Lucchesi}, {Lardo},
  {North}, {Irwin}, {Battaglia}, {Hill}, {Tolstoy}, {Venn}, {Helmi}, {Kaufer},
  {Primas}, \& {Shetrone}}]{theleretal2020}
{Theler}, R., {Jablonka}, P., {Lucchesi}, R., {et~al.} 2020,
  \bibinfo{title}{{The chemical evolution of the dwarf spheroidal galaxy
  Sextans},} \aap, 642, A176, \dodoi{10.1051/0004-6361/201937146}

\bibitem[{E. {Tolstoy} {et~al.}(2025){Tolstoy}, {Battaglia}, {Arroyo-Polonio},
  {Brown}, {van Essen}, {Massari}, {Sk{\'u}lad{\'o}ttir}, {Irwin}, {Taibi}, \&
  {Pritchard}}]{tolstoyetal2025}
{Tolstoy}, E., {Battaglia}, G., {Arroyo-Polonio}, J.~M., {et~al.} 2025,
  \bibinfo{title}{{A 3D view of dwarf galaxies with Gaia and VLT/FLAMES: II.
  The Sextans dwarf spheroidal},} \aap, 698, A53,
  \dodoi{10.1051/0004-6361/202554176}

\bibitem[{M.~G. {Walker} {et~al.}(2009){Walker}, {Mateo}, {Olszewski},
  {Pe{\~n}arrubia}, {Evans}, \& {Gilmore}}]{walkeretal2009}
{Walker}, M.~G., {Mateo}, M., {Olszewski}, E.~W., {et~al.} 2009,
  \bibinfo{title}{{A Universal Mass Profile for Dwarf Spheroidal Galaxies?},}
  \apj, 704, 1274, \dodoi{10.1088/0004-637X/704/2/1274}

\bibitem[{H. {Yang} {et~al.}(2025){Yang}, {Wang}, {Zhu}, {Li}, {Koposov},
  {Han}, {Li}, {Shi}, {Valluri}, {Riley}, {Dey}, {Rockosi}, {Palau}, {Aguilar},
  {Ahlen}, {Brooks}, {Claybaugh}, {Cooper}, {de la Macorra}, {Doel}, {Ferraro},
  {Forero-Romero}, {Gazta{\~n}aga}, {Gontcho A Gontcho}, {Gonzalez Morales},
  {Gutierrez}, {Guy}, {Honscheid}, {Ishak}, {Joyce}, {Kehoe}, {Kisner},
  {Kizhuprakkat}, {Kremin}, {Lahav}, {Landriau}, {Le Guillou}, {Medina},
  {Meisner}, {Miquel}, {Palanque-Delabrouille}, {Prada},
  {P{\'e}rez-R{\`a}fols}, {Rossi}, {Sanchez}, {Schlegel}, {Schubnell},
  {Silber}, {Sprayberry}, {Tarl{\'e}}, {Weaver}, {Zhou}, \&
  {Zou}}]{yangetal2025b}
{Yang}, H., {Wang}, W., {Zhu}, L., {et~al.} 2025, \bibinfo{title}{{The Dark
  Matter Content of Milky Way Dwarf Spheroidal Galaxies: Draco, Sextans, and
  Ursa Minor},} \apj, 993, 249, \dodoi{10.3847/1538-4357/ae07ce}

\bibitem[{S. {Yang} {et~al.}(2025){Yang}, {Zhang}, {Zhang}, \&
  {Li}}]{yangetal2025}
{Yang}, S., {Zhang}, S., {Zhang}, L., \& {Li}, H. 2025, \bibinfo{title}{{A
  Study of Abundance Patterns in the Sextans and Sculptor Dwarf Spheroidal
  Galaxies},} \apj, 981, 155, \dodoi{10.3847/1538-4357/adb57c}

\bibitem[{Y. {Zheng} {et~al.}(2025){Zheng}, {Yang}, {Zhang}, {Zheng}, {Wang},
  {Staveley-Smith}, {Tsai}, {Li}, {Liu}, {Hu}, {Chen}, {Quan}, {Zheng}, \&
  {Li}}]{zhengetal2025}
{Zheng}, Y., {Yang}, Y., {Zhang}, Y.-K., {et~al.} 2025,
  \bibinfo{title}{{PANCAKE: A Python-based Numerical Color─Magnitude Diagram
  Analysis Package},} \apjs, 279, 12, \dodoi{10.3847/1538-4365/add88c}

\end{thebibliography}
%\input{paper.bbl}

\begin{appendix}
\section{CMDs of Sextans sub-fields}

\begin{figure}
\includegraphics[width=9cm]{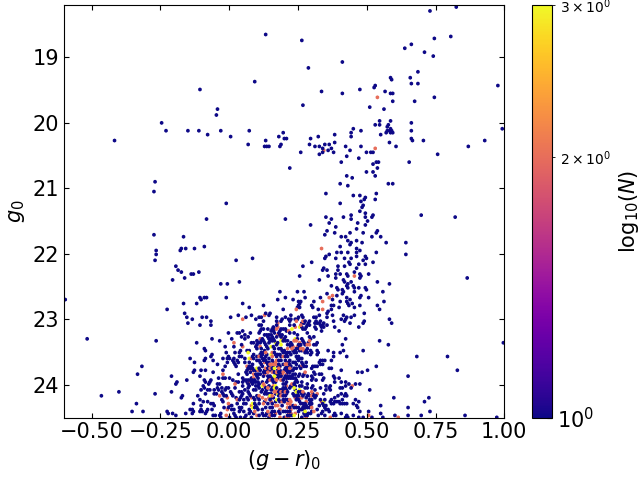}
\includegraphics[width=9cm]{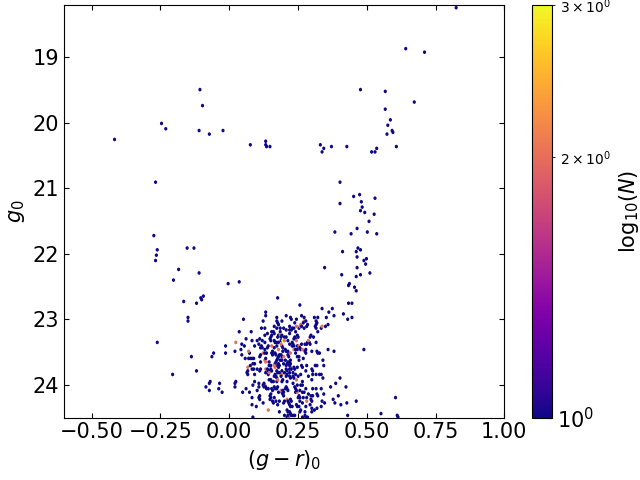}
\caption{Same as Figure~\ref{fig2} for the Sextans's core region.}
\label{fig2appendix}
\end{figure}

\begin{figure}
\includegraphics[width=9cm]{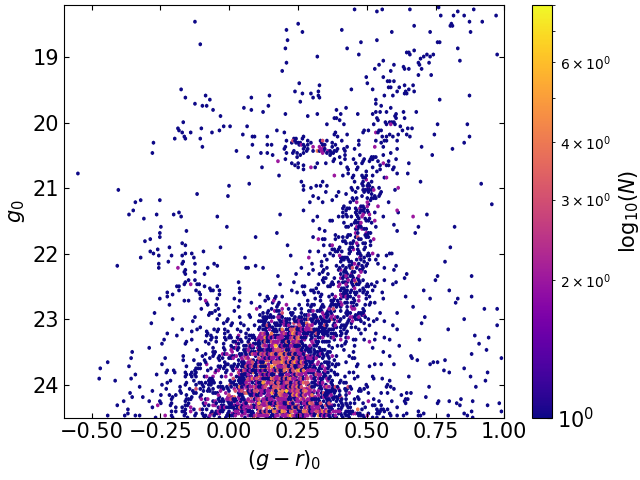}
\includegraphics[width=9cm]{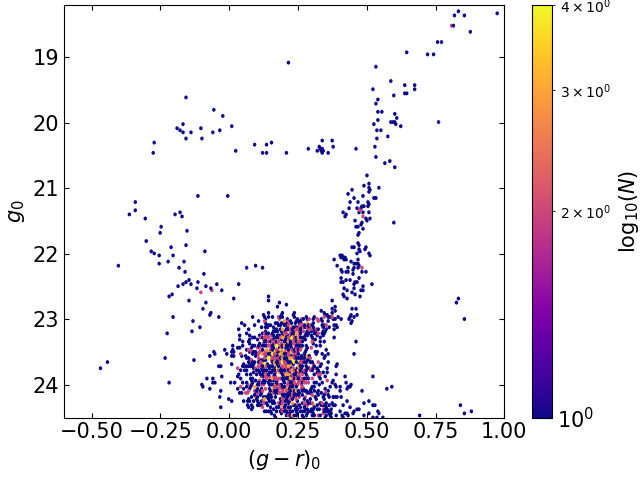}
\caption{Same as Figure~\ref{fig2} for the Sextans's ring region.}
\end{figure}

\begin{figure}
\includegraphics[width=9cm]{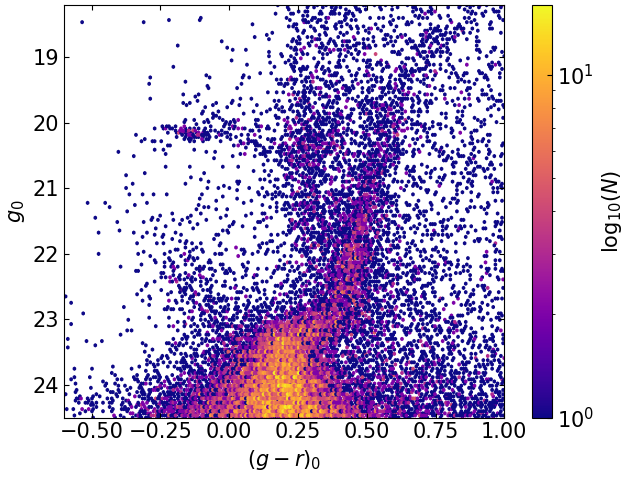}
\includegraphics[width=9cm]{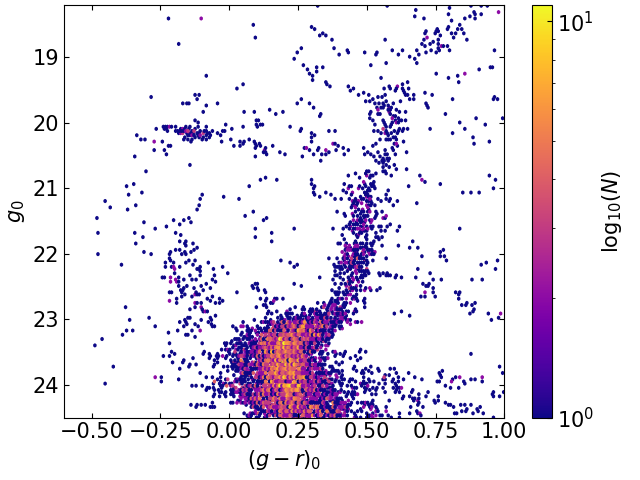}
\caption{Same as Figure~\ref{fig2} for the Sextans's outer body region.}
\end{figure}

\begin{figure}
\includegraphics[width=9cm]{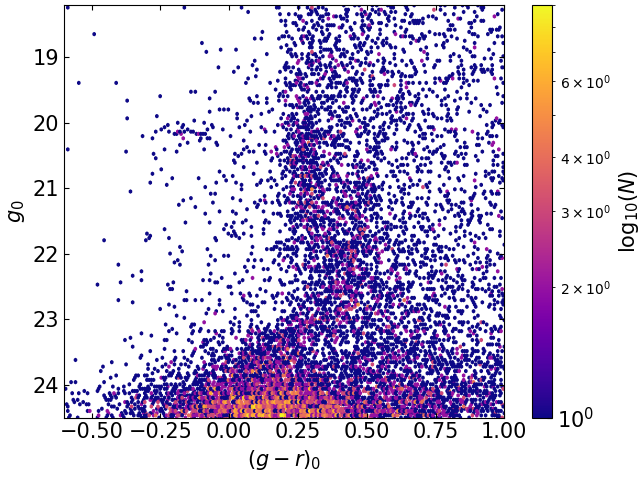}
\includegraphics[width=9cm]{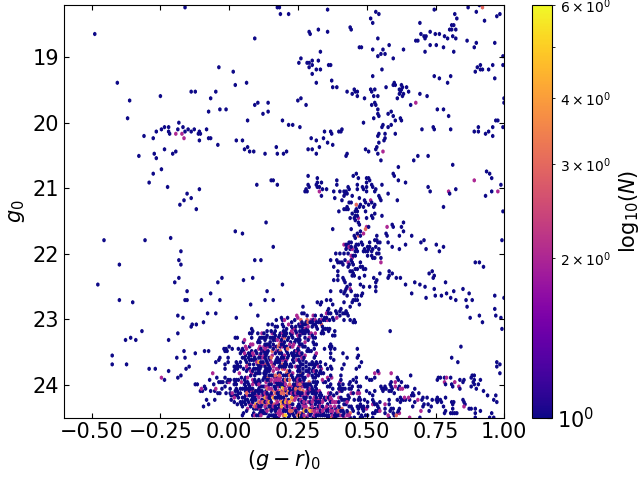}
\caption{Same as Figure~\ref{fig2} for the Sextans's outskirts region.}
\end{figure}

\begin{figure}
\includegraphics[width=9cm]{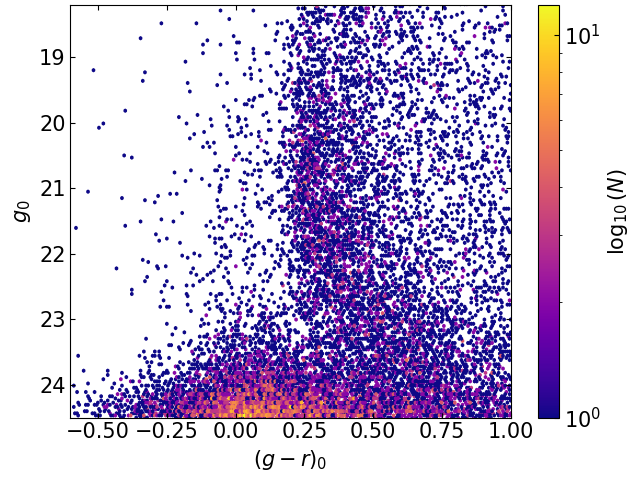}
\caption{Same as Figure~\ref{fig2} for the Sextans's reference region.}
\end{figure}

\begin{figure}
\includegraphics[width=9cm]{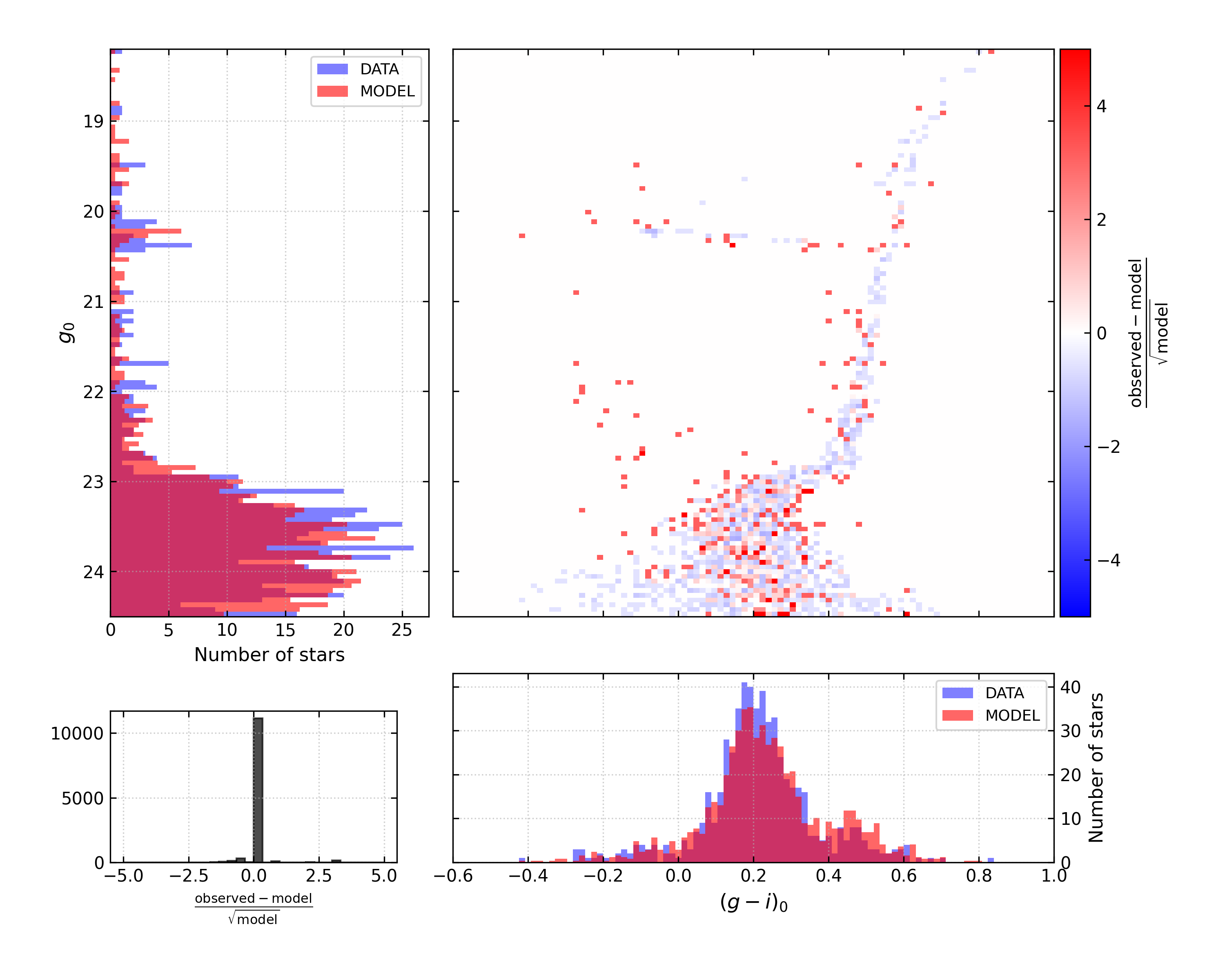}
\includegraphics[width=9cm]{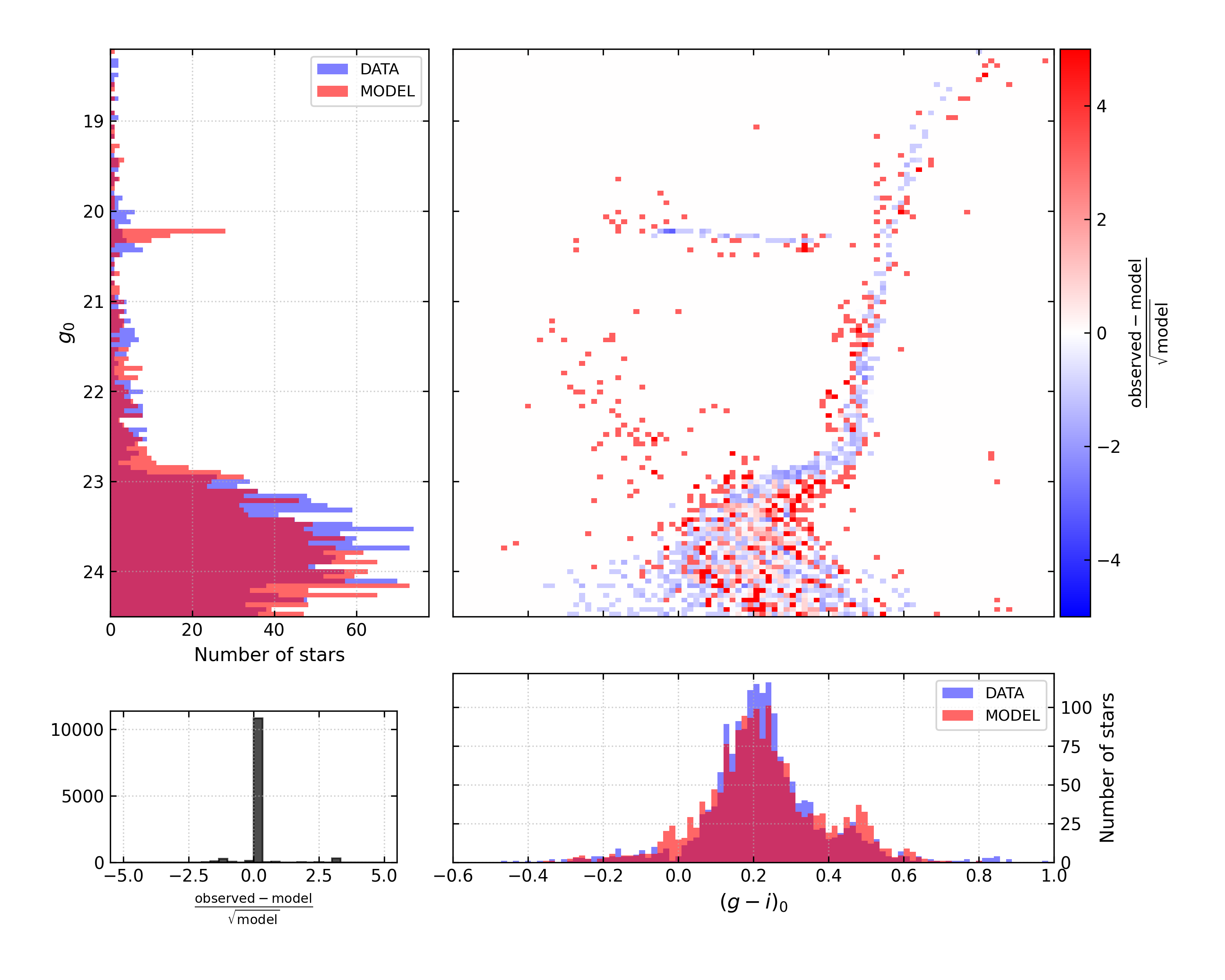}
\caption{Same as Figure~\ref{fig3} for the Sextans's core (left) and ring
(right) regions.}
\label{fig3appendix}
\end{figure}

\begin{figure}
\includegraphics[width=9cm]{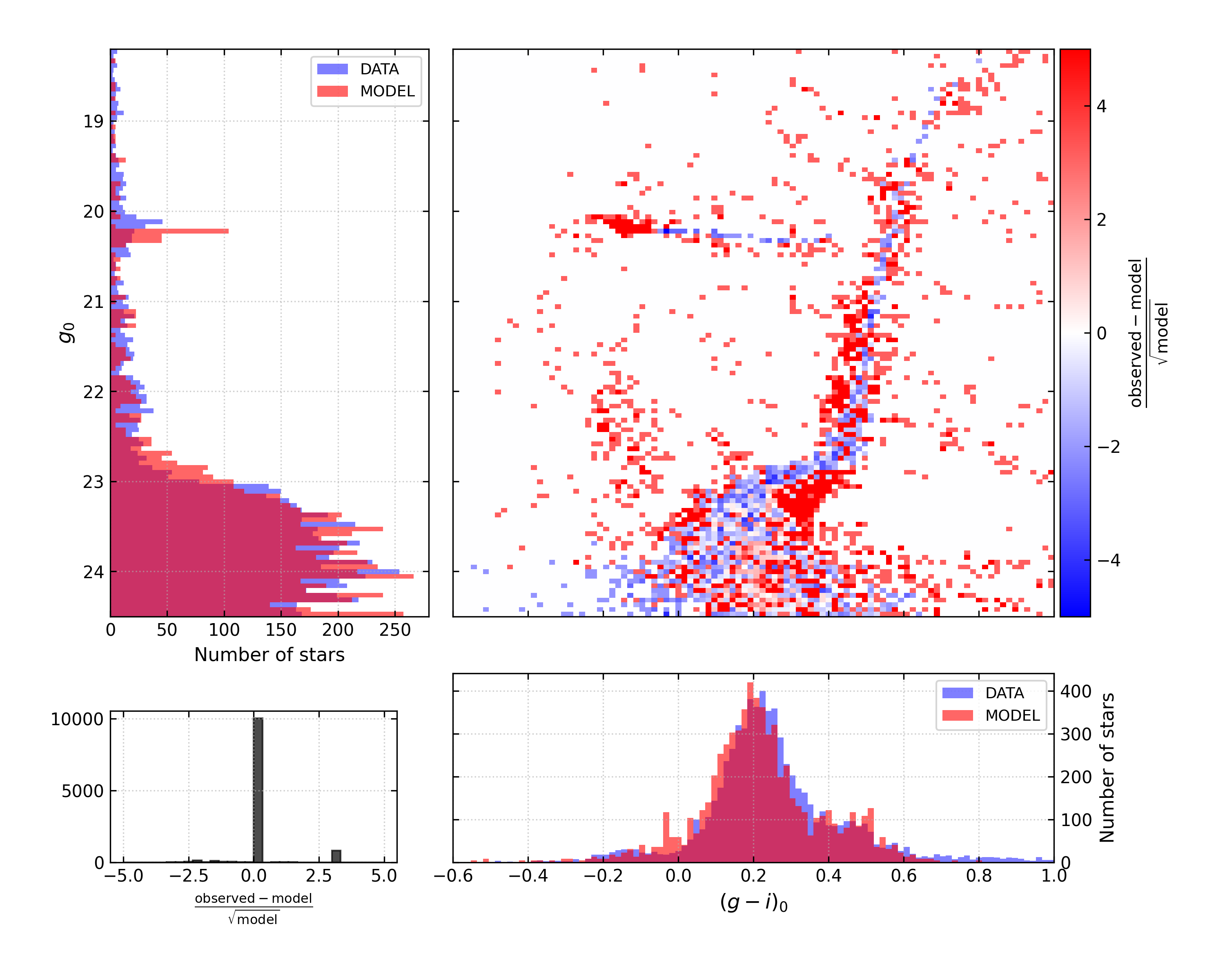}
\includegraphics[width=9cm]{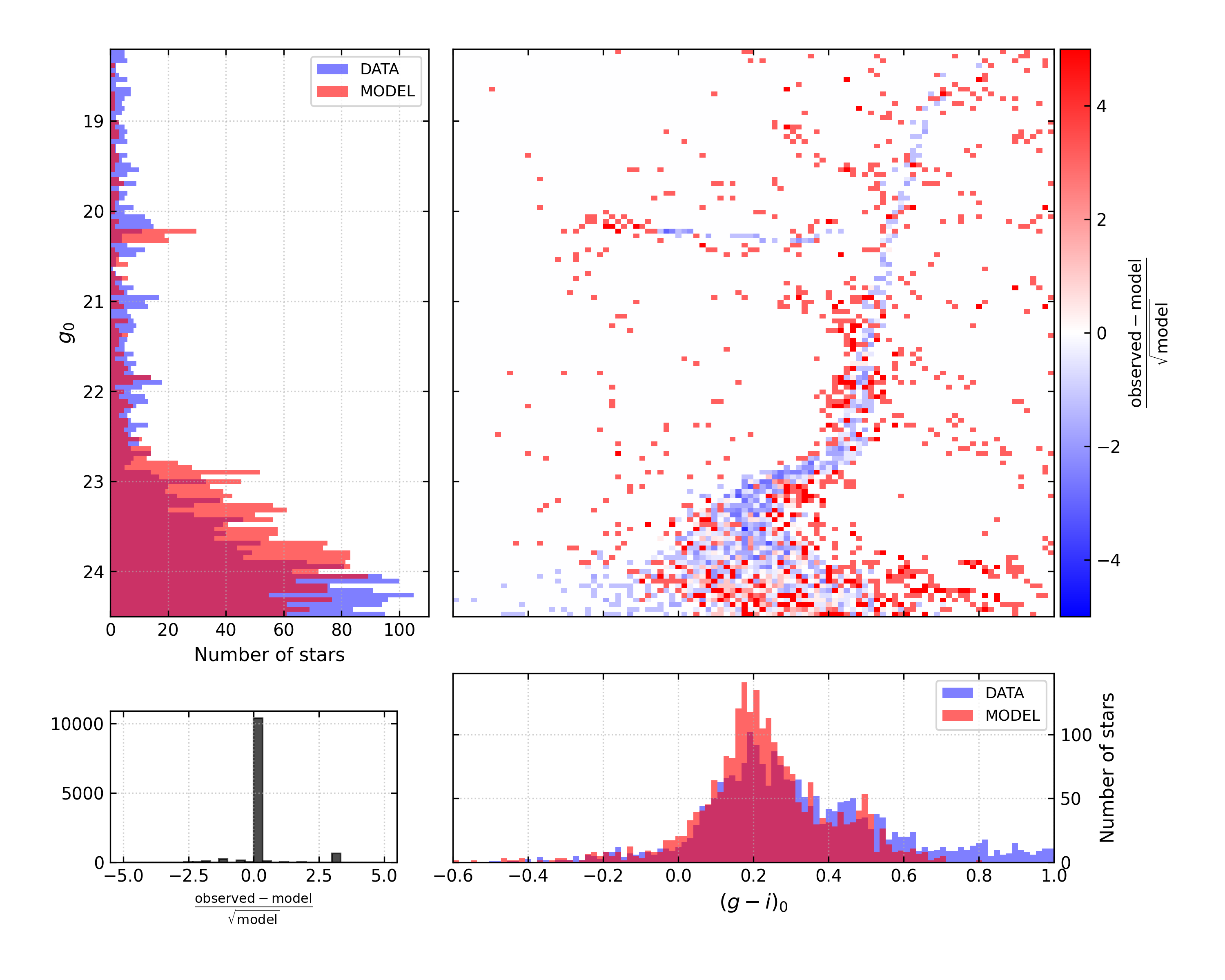}
\caption{Same as Figure~\ref{fig3} for the Sextans's outer body (left)
and outskirts (right) regions.}
\end{figure}

\end{appendix}

\end{document}